\documentclass[twocolumn,aps,floatfix]{revtex4-1}
\usepackage{amssymb}
\usepackage{amsmath}
\usepackage{graphicx}
\usepackage{pstricks}
\usepackage{ulem}
\usepackage{color}

\begin{document}

\title{The superfluid fountain effect in a Bose-Einstein condensate}

\author{Tomasz Karpiuk,$^{1,2}$ Beno\^{\i}t Gr\'{e}maud,$^{2,3,4}$ Christian Miniatura,$^{2,3,5,6}$ and Mariusz Gajda$^{7,8}$}                          

\affiliation{
\mbox{$^1$ Wydzia{\l} Fizyki, Uniwersytet w Bia{\l}ymstoku, 
            ulica Lipowa 41, 15-424 Bia{\l}ystok, Poland}  \\
\mbox{$^2$ Centre for Quantum Technologies, National University of Singapore, 3 Science Drive 2, Singapore 117543, Singapore} \\
\mbox{$^3$ Department of Physics, National University of Singapore, 2 Science Drive 3, Singapore 117542, Singapore} \\
\mbox{$^4$ Laboratoire Kastler Brossel, Ecole Normale Sup\'{e}rieure,} \\
\mbox{CNRS, UPMC; 4 Place Jussieu, 75005 Paris, France} \\
\mbox{$^5$ Institut Non Lin\'{e}aire de Nice, UMR 7335, UNS,} \\
\mbox{CNRS; 1361 route des Lucioles, 06560 Valbonne, France} \\
\mbox{$^6$ Institute of Advanced Studies, Nanyang Technological University, 60 Nanyang View, Singapore 639673, Singapore} \\
\mbox{$^7$ Instytut Fizyki PAN, Aleja Lotnik\'ow 32/46, 02-668 Warsaw, Poland} \\
\mbox{$^8$ Faculty of Mathemathics and Sciences, Cardinal Stefan Wyszy\'nski University, Warsaw, Poland} \\
}             

\date{\today}

\begin{abstract}

We consider a simple experimental setup, based on a harmonic confinement, where a Bose-Einstein condensate and a thermal cloud of weakly interacting alkali atoms are trapped in two different vessels connected by a narrow channel. Using the classical field approximation, 
we theoretically investigate the analog of the celebrated superfluid helium fountain effect. We show that this thermo-mechanical effect might indeed be observed in this system. By analyzing the dynamics of the system, we are able to identify the superfluid and normal components of the flow as well as to distinguish the condensate fraction from the superfluid component. We show that the superfluid component can easily flow from the colder vessel to the hotter one while the normal component is practically blocked in the latter. In the long-time limit, the superfluid component exhibits periodic oscillations reminiscent of the AC-Josephson effect obtained in superfluid weak links experiments.

\end{abstract}

\maketitle

\section{Introduction}

The experimental discovery of superfluidity in helium II by Kapitsa \cite{Kapitza} and Allen and Misener \cite{Allen} in 1938 has 
triggered a great theoretical interest in this phenomenon. One of the most spectacular effects related to superfluidity of 
helium II is its ability to flow through narrow channels with apparently zero viscosity. Extensive studies of this system were 
very important for the foundation of the theory of Bose and Fermi quantum liquids. In this system however, even at the lowest temperatures, the strong interactions between the helium atoms deplete the population of the Bose-Einstein condensate to about 10\% of the total mass whereas the superfluid fraction is almost 100\%.

The situation is substantially different with dilute ultracold atomic gases.
The first implementation of a Bose-Einstein condensation \cite{Cornell,Ketterle} in alkali atoms has opened new possibilities to 
explore Bose quantum liquids at much higher level of control. Indeed, contrary to liquid helium, large condensate fractions are routinely obtained with dilute atomic gases as the atoms are very weakly interacting. To date, many phenomena previously observed in liquid helium below the lambda point have found their experimental counterpart with ultracold alkali gases even if the experimental evidence of superfluidity in atomic condensates has been a very challenging task. One of the main signatures of superfluid flow is the generation of quantized vortices when the system is set into rotation. After many efforts such quantized vortices, and also arrays of vortices, were observed in 
atomic condensates \cite{Cornell_vortex, Dalibard,Ketterle_vortex}. Observation of the first sound \cite{Cornell-zero,Ketterle-zero}, of scissor modes \cite{Foot} or of the critical velocity \cite{critical-velocity} beyond which the superfluid flow breaks down, are other examples of the manifestation of this spectacular macroscopic quantum phenomenon in trapped ultracold atomic systems. 

In addition to the above-mentioned properties, helium II  exhibits also a very unusual feature related to the flow of heat. Variations of temperature propagate in this system in a form of waves known as the second sound. Both these extraordinary 
features, i.e.  non viscous flow and unusual heat transport, manifest themselves in full glory in the helium fountain effect, called  
also the thermo-mechanical effect.  Its first observation was reported by Allen and Jones \cite{Allen_Jones}. 
In their original setup, the lower part of a U-tube packed with fine emery powder was immersed into a vessel containing liquid helium II. A temperature gradient was created by shining a light beam on the powder which got heated due to light absorption. As a result of the temperature gradient, a superfluid flow is generated from the cold liquid helium reservoir to the hotter region. This flow can be so strong that a jet of helium is forced up through the vertical part of the U-tube to a height of several centimeters, hence the fountain effect name.

\begin{figure}[htb]
\resizebox{3.3in}{0.9in} {\includegraphics{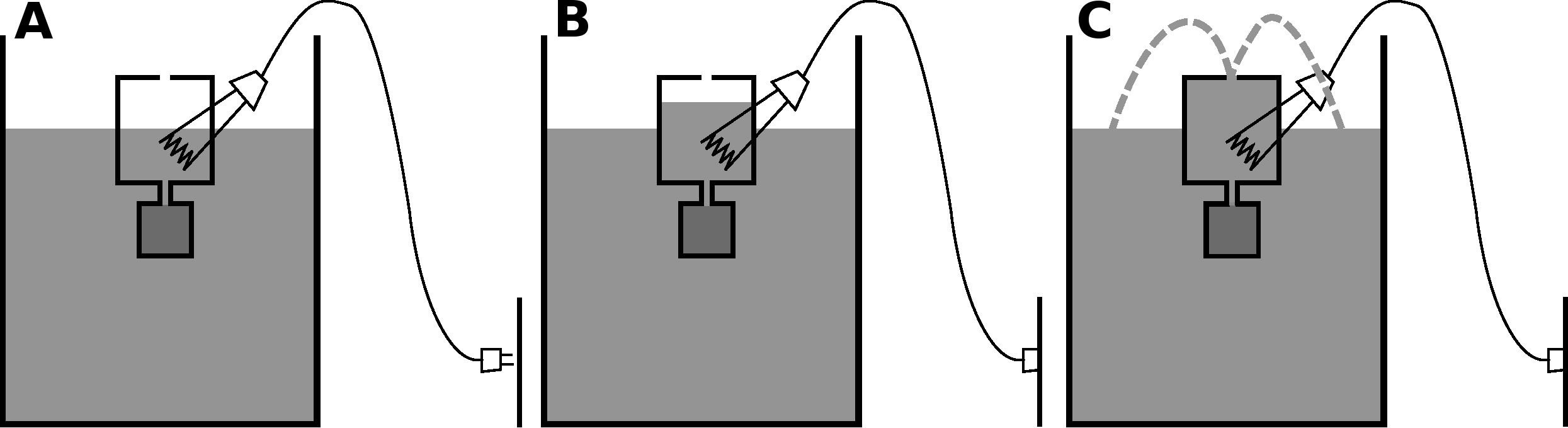}}
\caption{A cartoon picture showing the idea of the superfluid fountain experiment. A small vessel, connected to a bulb filled with emery powder forming a very fine capillary net, is immersed in a larger vessel containing liquid helium II. When the electric heater is off the superfluid liquid flows freely through the capillary net in the bulb and fills in the small vessel so that the liquid levels in both vessels are the same (panel A). When the superfluid helium inside the small vessel is heated, then the liquid level increases above the liquid level in the larger vessel (panel B). A continuous heating sustains the flow from the colder part of the system to the hotter one, an observation at variance with our ordinary everyday life experience. Eventually liquid helium reaches the top of the small vessel where it forms the helium fountain (panel C).
}
\label{fount}
\end{figure}

Up to now, there exists many different experimental implementations of this spectacular effect and one of them is shown in Fig.\ref{fount}. A small vessel, connected to a bulb filled with emery powder forming a very fine capillary net, is immersed in a larger container of liquid helium II. When the electric heater is off the superfluid liquid flows freely through the capillary net in the bulb and fills in the small vessel. As shown in panel (A) of Fig.\ref{fount}, the equilibrium is reached when the liquid levels in both vessels are the same. If now the superfluid helium inside the small vessel is heated then the level of the liquid in the smaller vessel increases above the level of the liquid in the big container, see panel (B). A continuous heating sustains the flow from the colder part of the system to the hotter one, an observation at variance with our ordinary everyday life experience. Eventually liquid helium reaches the top of the small vessel where it forms the helium fountain, see panel (C).

The explanation for this counter-intuitive thermo-mechanical effect is closely related to the notion of the second sound and to the two-fluid model developed by Tisza and Landau \cite{Tisza, Landau}. This approach assumes the existence of two co-existing components of the liquid helium: the superfluid 
and the normal one. The normal component is viscous and can transport heat. On contrary, the superfluid component has no viscosity and cannot transport heat. Because it is viscous, the normal component cannot flow through the capillary net but the superfluid can. Heat transport is thus forbidden because it can only be carried by the normal component. As a consequence, the system cannot 
reach thermal equilibrium and the temperature in the reservoir keeps smaller than the temperature in the small vessel. But heating of the superfluid component inside the small vessel leads to a reduction of the chemical potential in this vessel. In order to maintain thermodynamical equilibrium, this chemical potential drop has to be compensated by a superfluid flow from the reservoir. In other words, the equilibration of the chemical potentials in both vessels implies that the temperature difference between the two vessels is also accompanied by a pressure difference responsible for the fountain effect. 

The two-fluid model for helium II assumes a local thermal equilibrium which signifies a hydrodynamic regime where the collision time is the shortest time scale. If this is indeed the case for superfluid helium II, which is a strongly interacting system, it is generally not for trapped ultracold dilute atomic gases where reaching this regime proves extremely difficult. For example, second sound has only been observed recently \cite{second_sound}. As a consequence, the usual two-fluid model fails to apply.
Nevertheless, the existence of the fountain effect was suggested in \cite{Bagnato} on the basis of the hydrodynamic approach.  But we are not aware of any subsequent theoretical predictions about heat transport in weakly-interacting atomic condensates, or of any simulations of an effect similar to the helium fountain assuming a particular experimental arrangement.

The question of the nature of heat transport in these weakly-interacting atomic condensates seems to be well posed.  There are not many experiments where a non-equilibrium transfer of atoms related to temperature differences have been studied. We should recall here the experiment of the MIT group, where distillation of a condensate was observed \cite{Ketterle_distillation}. The authors studied how the superfluid system `discovers' the existence of a dynamically-created global minimum of the trapping 
potential and how the system gets to this minimum. Theoretical studies of the corresponding 1D situation suggested different dynamical behaviors of the thermal fraction and of the superfluid component which, in some sense, resemble the fountain effect \cite{Gorecka}.
Very recently, in analogy with electrical conductance in metals, particle transport through a mesoscopic channel between two macroscopic containers of particles (a source and a sink) was observed  with fermionic Li$^6$ atoms \cite{Esslinger}.
 
In the present work we theoretically study the non-equilibrium dynamics of a Bose-Einstein condensate which is driven by a temperature gradient. We will show that an effect qualitatively very similar to the helium fountain  may be observed in experiments with trapped ultracold dilute atomic gases.

The paper is organized as follows: in Sec. \ref{system}, we describe the system under consideration and our numerical procedure to prepare the initial state of the system and run its time evolution. To this end, we use the classical field approximation (CFA). Then in Sec. \ref{results} we present and analyze our numerical data. We show in particular that the thermo-mechanical effect is indeed present in our system and we highlight the question of distinguishing between the superfluid, normal, condensate and thermal components of the system. Finally, we give in Sec. \ref{concl} some concluding remarks and future work to address. To be self-contained, we present in the Appendix the CFA method used throughout this paper.

\section{Experimental system}
\label{system}

Following \cite{Phillips}, we consider here a cloud of Na atoms prepared in the $|3S_{1/2},F=1,m_F=-1\rangle$ state and confined in a 
harmonic trap with trapping frequencies $\omega_x = \omega_y = 2\pi\times 51$Hz and $\omega_z = 2\pi\times 25$Hz. 
The scattering length for this system is $a=2.75$nm. In subsequent calculations, we use the harmonic oscillator 
length $\ell_{{\rm osc}}= \sqrt{\hbar/m\omega_z} = 4.195\mu$m, time $\tau_{{\rm osc}} = 1/\omega_z = 6.366$ms and 
energy $\epsilon_{{\rm osc}} = \hbar\omega_z$ as space, time and energy units (oscillatory units).

\subsection{Preparation of initial states}

\begin{figure}[htb]
\resizebox{2.5in}{5.1in} {\includegraphics{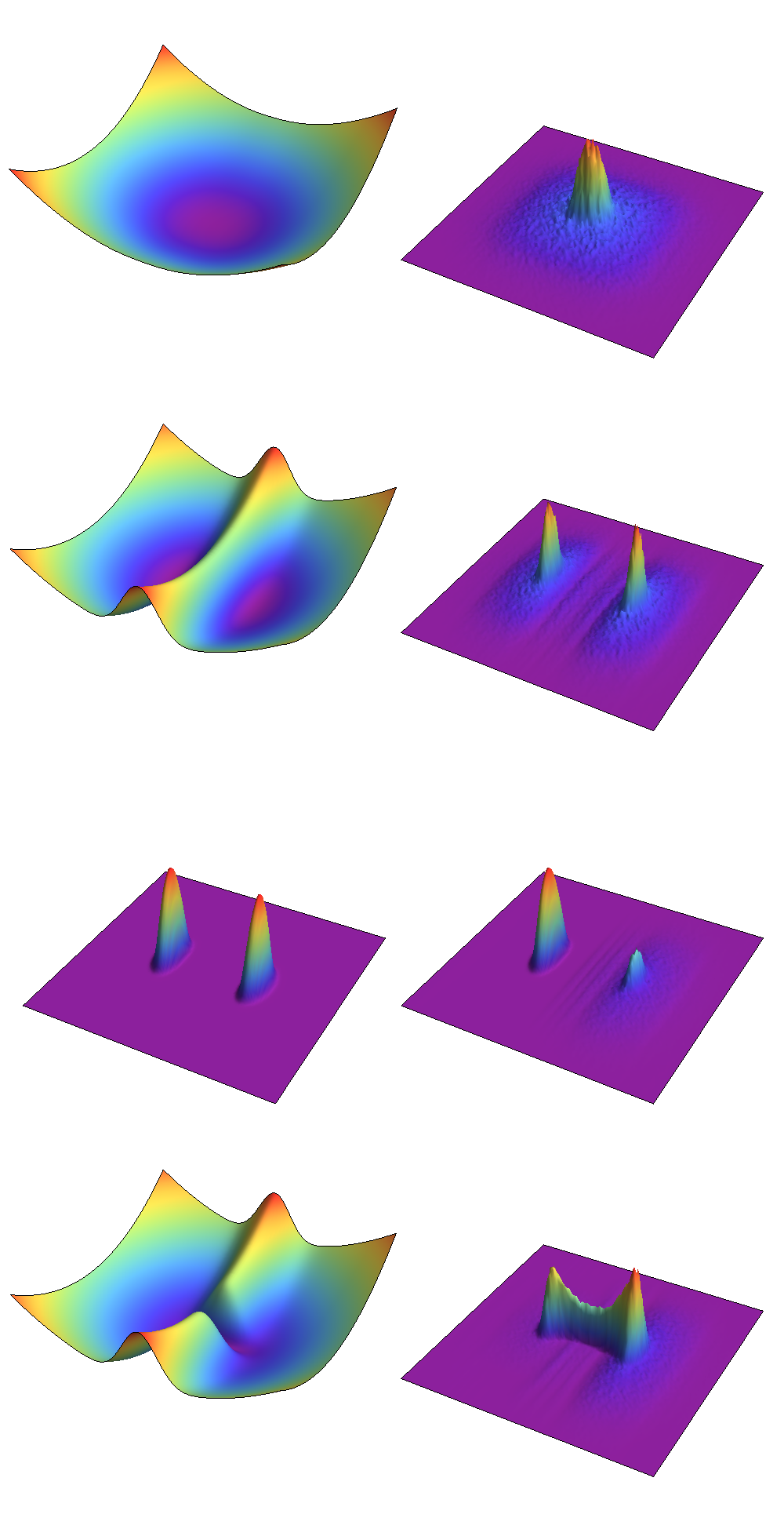}}
\caption{(Color online). Three-dimensional surface plots of the trapping potential and averaged atomic column density at the different stages of the simulations. The upper row shows the initial harmonic trap (left) during the preparation of the initial state. The corresponding atomic density at thermal equilibrium is shown on the right side. 
The second row shows the double-well trap obtained by rising the barrier at the center of the harmonic trap (left) and the corresponding equilibrium atomic density (right). 
The third row shows the density of atoms
in the double-well trap at zero temperature (left) and when the temperatures in each well are different (right). In this example, the left well contains a pure condensate ($T=0$) whereas the condensate fraction in the right well is $20$\% ($T=100$nK). The last row shows the two wells connected through 
a thin channel (left). The corresponding atomic density at some stage of the evolution is shown on the right.}
\label{traps}
\end{figure}

The preparation of an initial state in the harmonic trap $V_{tr}({\bf r}) = \frac{1}{2} m (\omega_x^2 x^2 +\omega_y^2 y^2 + \omega_z^2 z^2)$ 
follows the CFA steps described in the Appendix. An example of such state is shown in the first row of Fig.\ref{traps}. In this particular 
case the temperature
of the system is $100$nK and the condensate fraction is about $20$\%. We also prepared two more initial states corresponding to different 
condensate fractions ($50$\% and $100$ \%) i.e.  temperatures $T$, see Table \ref{table1}.
\begin{table}[htb]
 \begin{tabular}{|c|c|c|c|c|}
\hline
  $N$ & $N_0/N$ & $T [nK]$ & $k_B T [\epsilon_{{\rm osc}}]$ & $\mu [\epsilon_{{\rm osc}}]$ \\
\hline \hline
250000 & 1.0 & 0.	& 0.	& 22.7 \\
250000 & 0.5 & 84.	& 69.1	& 16.2 \\
250000 & 0.2 & 100.	& 83.7	& 12. \\
\hline
 \end{tabular} 
\caption{ Numerical values of the condensate fraction $N_0/N$, temperature $T$, thermal energy $k_B T$, and chemical potential $\mu$ used in our simulations.}
\label{table1}
\end{table}
When $T=0$, the initial state is simply the ground state of the Gross-Pitaevskii equation.

\begin{figure}[htb]
\resizebox{3.2in}{2.in} {\includegraphics{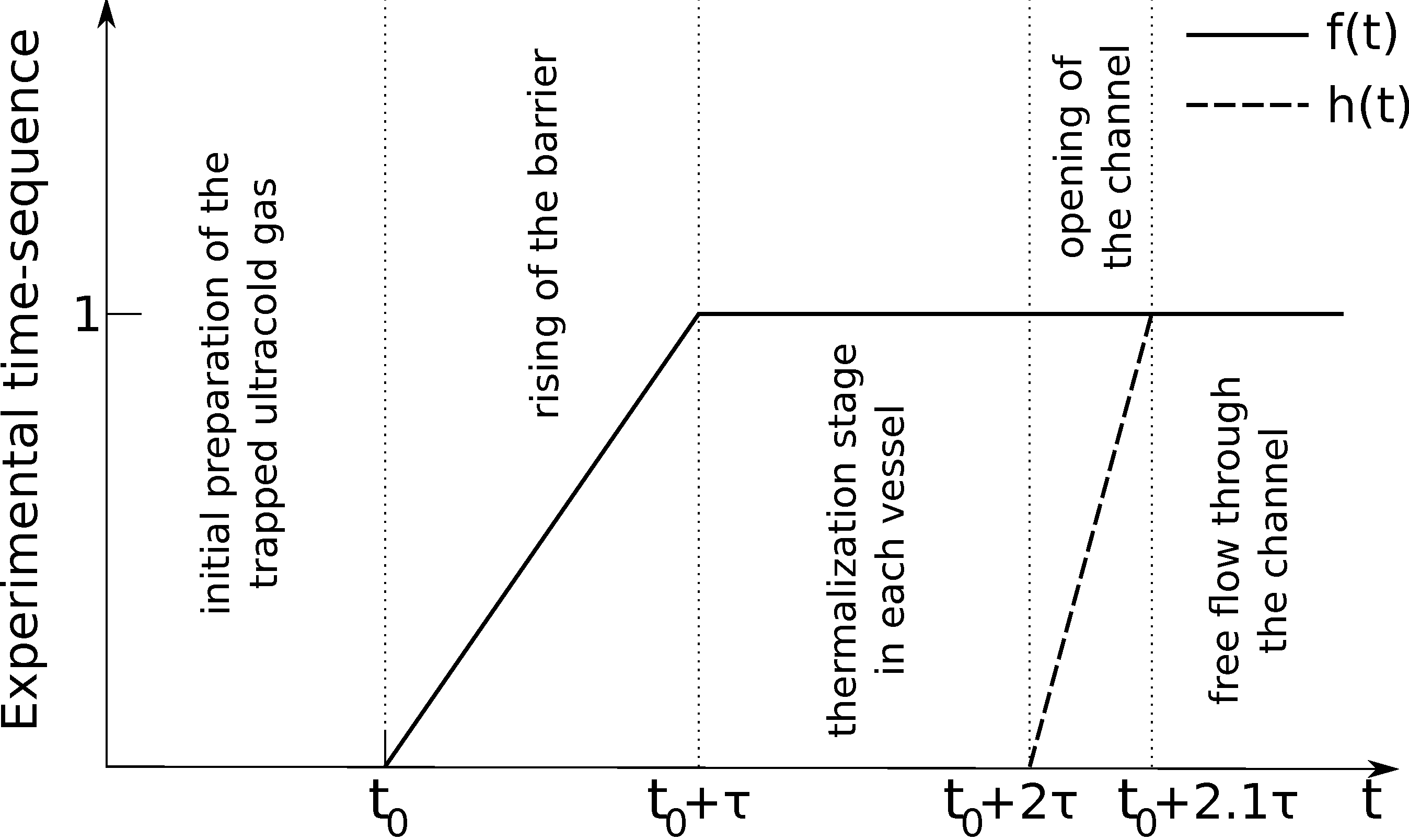}}
\caption{Sketch of the experimental time-sequence. Solid and dashed lines correspond to linear time-ramps $f(t)$ and $h(t)$ respectively.}
\label{time}
\end{figure}

In the next step, we split the cloud of atoms into two approximately equal parts by rising a Gaussian potential barrier $V_b({\bf r},t)= V_b({\bf r})  f(t)$ at the center of the harmonic trap by means
of a linear time-ramp $f(t)$, see Fig.\ref{time}. Such a barrier, with height $V_b$ and width $w_b$
\begin{equation}
 V_b({\bf r}) = V_b \, e^{-x^2/w_b^2},
\end{equation}
can be created by optical means using a blue-detuned laser light sheet perpendicular to the x-direction.
The barrier is ramped at a time $t_0$, chosen at the end of the initial equilibration phase, and we have fixed the barrier rising time at $\tau=78.54 \tau_{{\rm osc}}$ in our simulations. After this perturbation, we let the system reach again equilibrium in 
the double-well trap by evolving the state for an additional time $\tau$. Finally the system is split into two separate clouds containing each about $125000$ atoms. 
In all our simulations, the barrier parameters are fixed at $V_b = 432 \epsilon_{{\rm osc}}$ and
 $w_b = 2.529 \ell_{{\rm osc}}$ ($\approx 10.6\mu$m). The full width at half-maximum (FWHM) of the barrier is $W_b = 2 \sqrt{\ln{2}} w_b= 4.21\ell_{{\rm osc}}$ ($\approx 17.7\mu$m). The height of the barrier has been chosen much larger than $k_B T$ and the chemical potential $\mu$ so that both thermal and condensed atoms cannot flow through the barrier, see Table \ref{table1}.

As we can create the equilibrium state in a double-well potential corresponding to different initial temperatures, we can also easily prepare 
our system in a state where temperatures in both wells are different. This can be done by replacing the zero temperature component in 
the right well by a nonzero temperature cloud as shown in the third row of Fig.\ref{traps}. The numbers of atoms in each well are 
approximately equal. We have designed all steps of the preparation stage of the initial state of two subsystems with different temperatures having in mind a possible and realistic experimental realization. Only the last step, i.e. replacing the zero temperature component in one subsystem by 
a finite temperature state has to be done differently in the experiment. Heating only one subsystem localized in a given well could be done by a temporal modulation of the well, followed by a thermalization. 

\subsection{Opening the channel between the two vessels}

Having prepared two subsystems at different temperatures separated by the potential barrier, we can now study their dynamics when a thin channel is rapidly opened between the two wells. This is done by switching on the channel potential $V_c({\bf r},t) = V_c({\bf r}) h(t)$, where the linear time-ramp $h(t)$ starts after the equilibration of the two subsystems created by the barrier, i.e. at time $t_0+2\tau$, see Fig.\ref{time}. Its duration has been fixed to $\tau/10$ in all our numerical simulations.

When the channel is opened the two clouds of atoms come into contact. Since we start with different condensate fractions in both traps, there will be a chemical potential mismatch between the two clouds. In turn, both wave functions evolve differently and there will be a random phase step where the wave functions touch. We numerically found that this phase step can drive the flow of about $3$\% of the atoms for the considered geometry. Considering that the initial condensate fractions in both vessels differ at least by $50$\%, the effect of this initial relative random phase is barely visible in our numerical results and can be discarded. However it is worth mentioning that this phase mismatch may be crucial when the condensate fractions in the two vessels are comparable.

\begin{figure}[htb]
\resizebox{3.4in}{1.7in} {\includegraphics{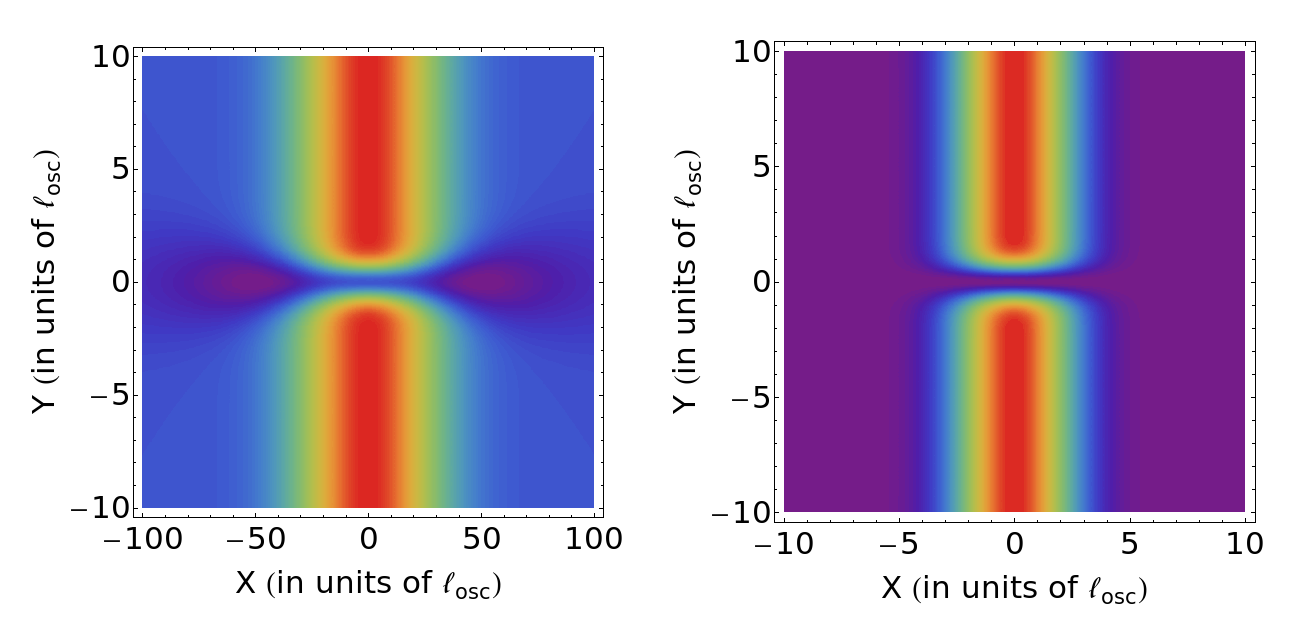}}
\caption{(Color online). Comparison between the combined barrier and channel potential obtained by using a focused Gaussian laser beam (left frame) and the one used in our simulations (right frame). In the first, atoms get trapped in the "potholes" and the number of atoms in the vessels has to be increased in order to observe the fountain effect.
}
\label{potholes}
\end{figure}

From an experimental point of view, there are various ways to create the channel potential $V_c({\bf r})$. For example, starting from an harmonic trap, one could 
use two orthogonal sheets of blue-detuned laser light propagating in the $(Oy,Oz)$ plane.
These two sheets build together the barrier described earlier in this section and by putting two obstacles along their direction of propagation, one would create two shadows. Their intersection would open the desired channel between the two wells but the minimum channel width would then be constrained by the diffraction effects induced by the two obstacles. However widths of the order of few $\mu$m should be feasible. Alternative methods would be to use TE$_{0,1}$ Hermite-Gaussian modes, or properly designed separate traps \cite{Foot2,Tiecke, Chip1,Chip2} and then focus a red-detuned Gaussian beam.
The corresponding channel potential would be:
\begin{eqnarray}
 V_g({\bf r}) &=& -V_b \frac{w_c^2}{w_c^2(x)} \, e^{-\frac{(y^2+z^2)}{w_c^2(x)}} \, ,\\
 w_c(x) &=& w_c \sqrt{1+x^2/w_b^2} \, ,\nonumber
\end{eqnarray}
where the Rayleigh length $x_R= k_Lw_c^2$ of the channel laser beam ($k_L$ is the laser wavenumber) has been matched to the barrier parameter $w_b$. For 
$w_c = 3.5\mu$m, one would have $w_b = 133\mu$m. The sum of the barrier potential $V_b({\bf r})$ and of the new channel potential $V_g({\bf r})$ is shown in the left frame of Fig.\ref{potholes}. In this case, the opened channel would have two "potholes" separated by a relatively small barrier and these spurious wells would trap atoms. In order to observe a superfluid flow and the fountain effect, one would then have to make sure that the chemical potential $\mu$ is larger than this small barrier height $\approx V_b/5$. We have run numerical simulations (not shown here) and checked that the fountain effect is indeed present in this case.

As this spurious trapping would unnecessarily complicate (but not kill) our proof-of-principle analysis of the fountain effect, we have chosen to work with the following channel potential in all our numerical simulations:
\begin{equation}
 V_c({\bf r}) = -V_b \, e^{-(y^2+z^2)/w_c^2}\ e^{-x^2/w_b^2}.
 \label{channelpot}
\end{equation}
It has the opposite barrier strength $V_b$, a Gaussian profile with width $w_c$ in the ($Oy,Oz)$ plane and same width $w_b$ as the barrier potential along $Ox$. The sum of $V_b({\bf r})$ and $V_c({\bf r})$ creates a smooth channel between the two vessels as it is shown in the right frame of Fig.\ref{potholes}. 

The FWHM-width of the channel is $W_c=2 \sqrt{\ln{2}} w_c$. The final shape of the total potential (harmonic trap included) is shown in the last row of Fig.\ref{traps} on the left, while a typical example
of the column density of the evolving atomic cloud is shown on the right. In our subsequent numerical simulations we will use different channel widths $W_c$ to compare the behavior of the thermal flow to the superfluid one. 


At this point, as evidenced by the right panel in Fig.\ref{potholes}, we would like to highlight the similarity between our trap design and the geometry used in superfluid Josephson weak links experiments where two superconductors or two superfluids are coupled to each other \cite{WeakLinks}. For superconductors, weak links are realized through tunnel junctions. For superfluids, one connects two reservoirs by an aperture junction with a size of the order of the healing length of the superfluids, a situation similar to ours. In weak links experiments, the two macroscopic wavefunctions leak into each other and couple, giving rise to an AC-Josephson current associated with a constant chemical potential difference between the two containers. This oscillating current violates our classical intuition that a pressure head applied across the fluid in a hole should result in unidirectional flow. We'll see later that our simulations do evidence a similar sine-like current (see Section \ref{SN}).

\section{Numerical Results}
\label{results}

\begin{figure}[htb]
\resizebox{3.in}{5.81in} {\includegraphics{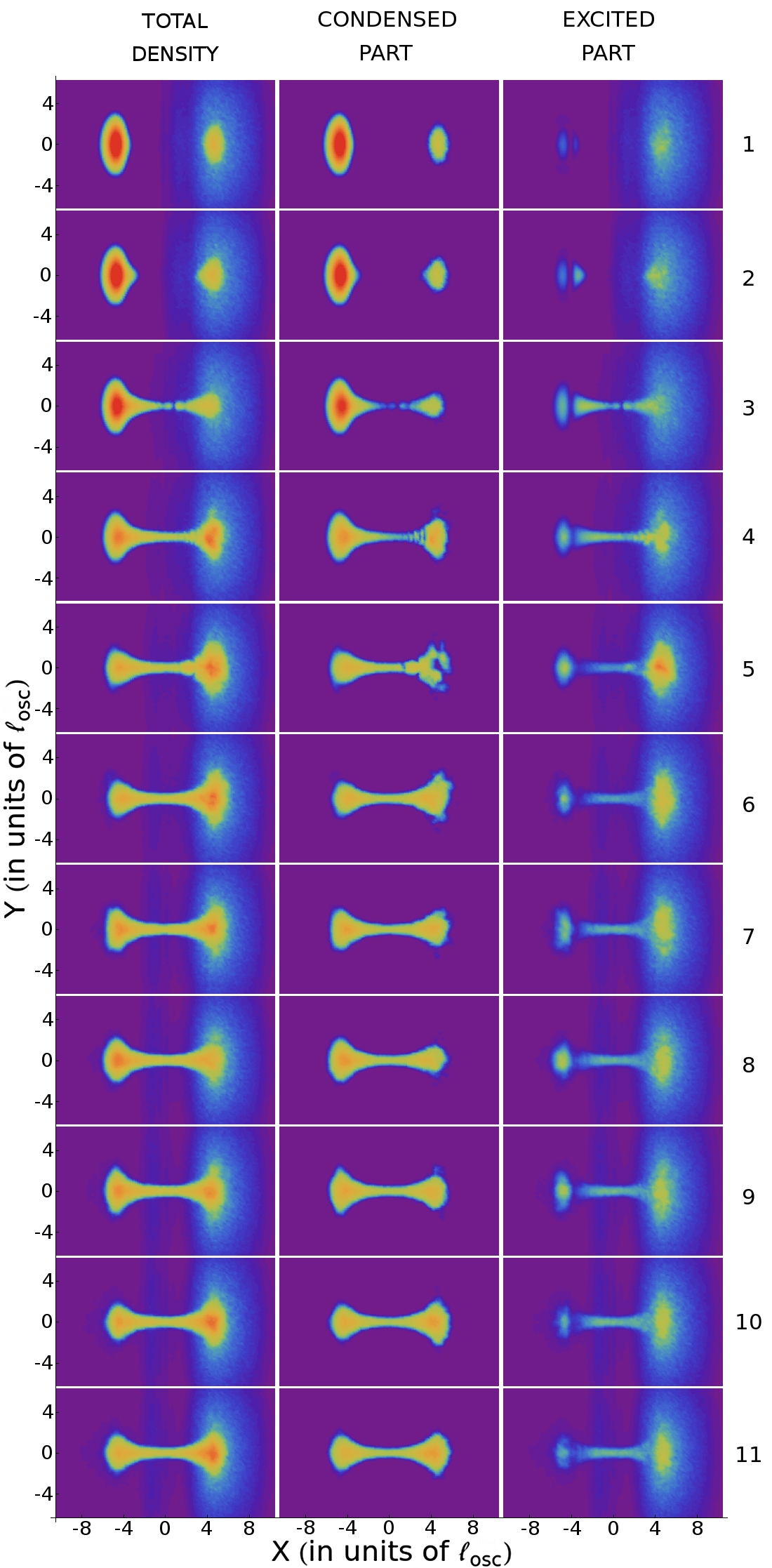}}
\caption{(Color online). Snapshots of the time evolution of the column atomic densities when the cold left vessel (pure condensate, $T=0$) and the hot right vessel (condensate fraction $20$\%, $T=100$nK) are connected by a channel. The initial number of atoms in each vessel is about $125000$. The left column of the different frames shows the total atomic density, the middle column shows the condensate density and the right column shows the density of thermal atoms. The channel width is  $W_c = 2.4 \ell_{{\rm osc}}$ ($10\mu$m). The time interval between the different frames is about $2.5 \tau_{{\rm osc}}$ ($\approx 15.9$ms).}
\label{denpo}
\end{figure}

The main observations of this paper concern the time evolution of two dilute atomic clouds at two different temperatures and initially prepared in two different potential wells (vessels). At a certain time, a "trench" is dug in the potential barrier separating the two vessels and the atoms can flow from one vessel to the other through the channel which has been opened. For classical systems one would expect a heat transport from  
the hotter cloud to the colder one, followed by a fast thermalization process. The hot vessel is the potential well on the right and it contains only $20$\% of condensed atoms ($T=100$nK). The left well is the cold vessel and it initially contains a pure condensate ($T=0$). In our simulations, we clearly see that, shortly after the two vessels are connected, the condensate is flowing very fast from the left cold vessel to the right hot vessel as shown in Fig.\ref{denpo}. 
In the six initial frames we clearly see that the atomic density in the right hot vessel is increasing significantly while it is decreasing in the left cold vessel. During the same time there is no visible transfer of thermal atoms from the hot vessel to the cold one. Atoms from the cold vessel are rapidly injected into the hot vessel. This scenario clearly has the flavor of the helium fountain experiment where the superfluid helium is flowing from the colder big vessel to the smaller hot vessel through a thin net of capillaries and finally streams through the small hole in the lid to form a jet. In our case we do not see a true fountain effect but instead some increase of the atomic density in the hot vessel. 
In fact this physical effect could be easily observed in an experiment using standard imaging techniques.

One has to note that, in the original helium fountain experiment, there is always a very big reservoir of superfluid atoms. Therefore the fountain effect can persist as long as the small vessel is heated. In our case the initial number of atoms in each wells is the same. The reservoir of cold atoms is thus almost emptied very fast. Then the situation gets reversed: the right vessel contains more cold atoms than the left one and the atomic cloud starts to oscillate back and forth between the two vessels. This is seen in Fig.\ref{denpo}, where frames $6-11$ show temporal oscillations of the total atomic density between the two vessels (left column).

\subsection{Condensate and thermal component}

The above qualitative findings can be justified quantitatively. To this end we first have to split the classical field into condensed and thermal components 
as described in the Appendix. The evolution of these components is shown in the middle and the right panels of Fig.\ref{denpo}. 
The flow starts when the channel between the two vessels is fully opened, which approximately corresponds to the third frame in Fig.\ref{denpo}. Analyzing the condensate part, we see that its initial flow is quite turbulent and a series of shock waves appears (frames 3 -- 5). 
This happens for two reasons. First, as previously mentioned, there is a random phase jump where the two clouds touch resulting in the creation of one or two gray solitons. Second, atoms flowing fast from the left vessel to the right one are reflected back by the boundaries of the right vessel and try to flow again to the left vessel.
As a result thermal atoms are produced in 
the right well (frame 5, left column) and the condensate gets fragmented (frame 5, middle column). After this initial turbulent evolution,  the flow becomes laminar. We have checked that these initial effects are significantly reduced when the temperature difference between the two subsystems is smaller.
 
A quantitative analysis of the dynamics requires an estimation of temperature of both subsystems. In this dynamical nonequilibrium situation, the notion of temperature is questionable. However we can use the condensate fraction in the left and the right well as an estimate of the 'temperature' of both subsystems. To this end, using Eq.\eqref{funk}, we split the relative occupation numbers of the one-particle density matrix modes into left and right components:
\begin{eqnarray}
&n_k^L(t)=\int_{-\infty}^0 dx \int_{-\infty}^\infty dy \;  \bar{\varrho}_k(x,y,x,y;t)\, , \nonumber \\
&n_k^R(t)=\int_{0}^\infty dx \int_{-\infty}^\infty dy \; \bar{\varrho}_k(x,y,x,y;t) \, .
\end{eqnarray}
This gives, for each vessel, the condensate, the thermal cloud and the total relative occupation numbers: 
\begin{eqnarray}
&n_0^{X}(t) \, , \nonumber \\
&n_T^{X}(t) = \sum_{k=1}^{K} n_k^{X}(t) \, , \nonumber \\
&n_{X}(t) = n_0^{X}(t)+n_T^{X}(t) \, ,
\label{relocpeq}
\end{eqnarray}
where $X=L,R$.

\begin{figure}[htb]
\resizebox{3.4in}{2.1in} {\includegraphics{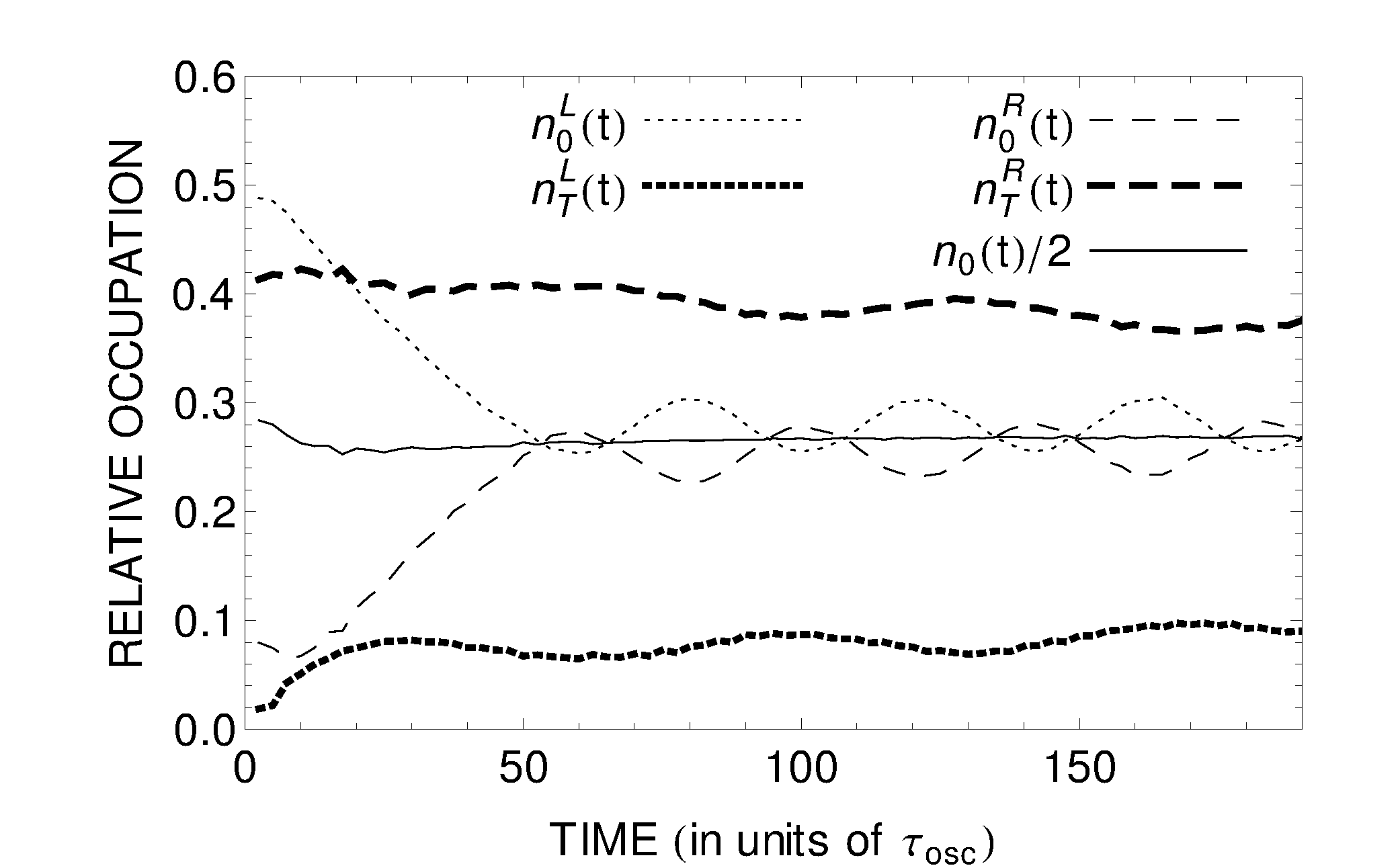}}
\resizebox{3.4in}{2.1in} {\includegraphics{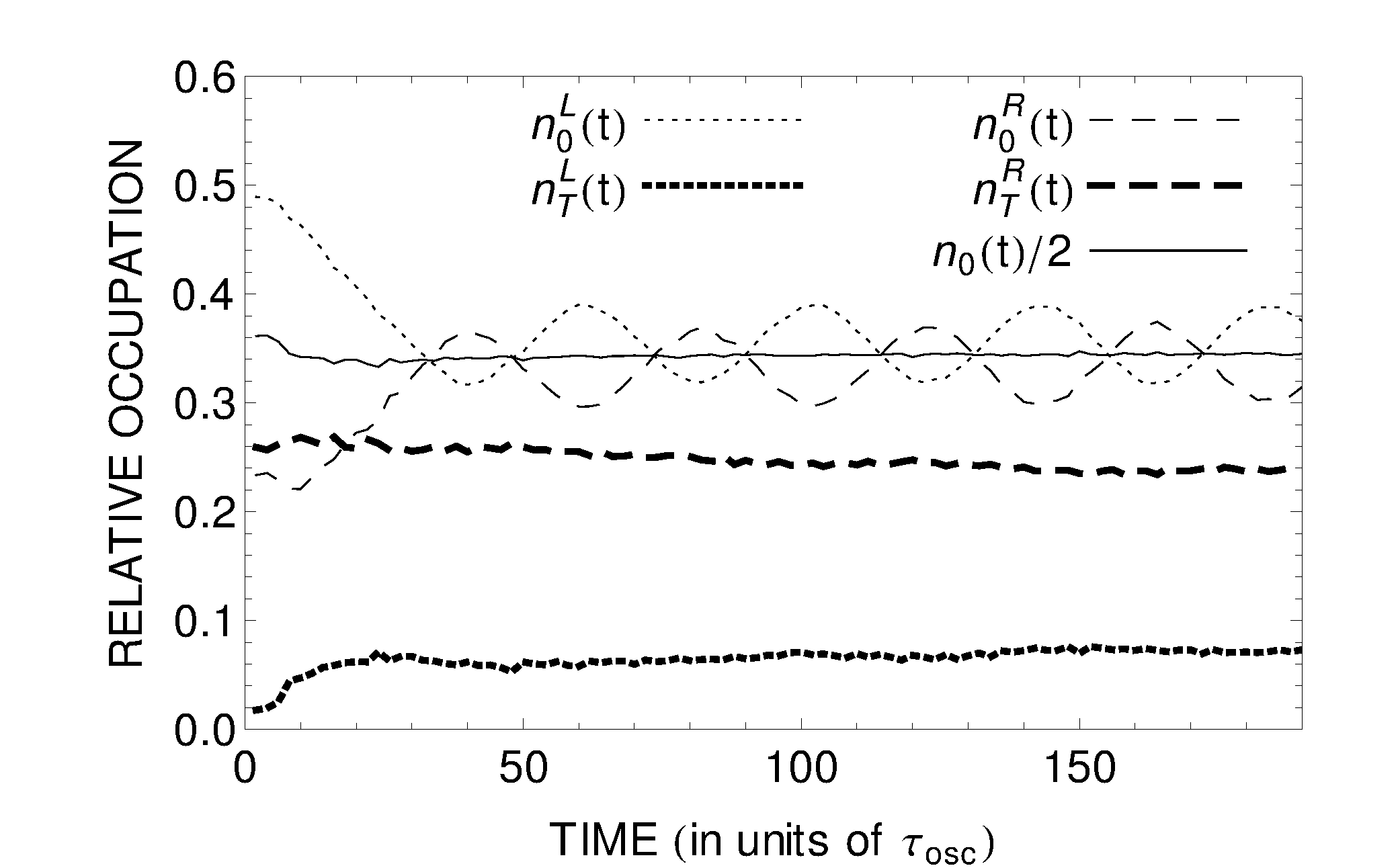}}
\caption{
Time evolution of the condensate and thermal relative occupation numbers, as given by Eq.\eqref{relocpeq}, in the left and right vessels for two different initial condensate fractions in the right well. The time unit is $\tau_{{\rm osc}} = 6.366$ms. Top frame: initial right condensate fraction of 
 $20$\% ($T=100$nK), final channel width $W_c=0.96\ell_{{\rm osc}}$ ($\approx 4\mu$m). Bottom frame: initial right condensate fraction of $50$\% ($T=83$nK), final channel width $W_c=\ell_{{\rm osc}}$ ($\approx 4.2\mu$m).
Condensate relative occupation numbers: $n^L_0(t)$ (thin dotted line) and $n^R_0(t)$ (thin dashed line). Thermal relative occupation numbers: $n_T^L(t)$ (thick dotted line) and $n_T^R(t)$ (thick dashed line). As one can see, after some time, the left and right condensate relative occupation numbers oscillate around half the total condensate fraction $n_0(t)/2$ (thin solid line) whereas the thermal fractions stay roughly constant.
}
\label{relocppo1}
\end{figure}

We have drawn the above quantities in Fig.\ref{relocppo1} for two different initial condensate fractions in the right well, $20$\% (T=$100$nK) for the top frame and $50$\% for the bottom frame ($T=84$nK). The thin and thick lines correspond 
to the condensate and thermal fractions respectively. The main observations are the following: (i) The initial injection of the left condensate at $T=0$ into the right well lasts about $47 \tau_{{\rm osc}}$ ($300$ms) in the upper frame, and about $31 \tau_{{\rm osc}}$ ($200$ms) in the lower frame; (ii) After the initial injection, the condensate fractions 
in both wells oscillate with a small amplitude around a mean value  -- some condensed atoms flow from one well to the other; (iii) The thermal components stay almost constant in both wells.   

However, a more detailed analysis shows some initial increase of the thermal component during the first $8-16 \tau_{{\rm osc}}$ ($50-100$ms) in the left well which is followed 
by a very slow flow of the thermal cloud from the hot to the cold part of the system. The initial increase of the thermal component can 
be easily explained. First of all, the opening of the channel between the two wells is not adiabatic and a thermal fraction is excited in the process -- see the first three panels in Fig.\ref{denpo}. Secondly, the initial flow of the condensed component is very fast and 
turbulent so it is another source of thermal excitations. Finally a small thermal fraction of atoms is initially present in the region of the barrier. These atoms form a strip along $Oy$ and perform oscillations with a small amplitude inside the barrier which is visible in the thermal components of the top frame of Fig.\ref{relocppo1}. This effect is reduced by lowering the initial temperature as shown in the bottom frame where the modulation of the thermal components is hardly noticeable. In fact there are still about $2-3$\% atoms in the barrier resulting in about $1$\% modulation of the thermal fraction.

\begin{figure}[htb]
\resizebox{3.2in}{2.1in} {\includegraphics{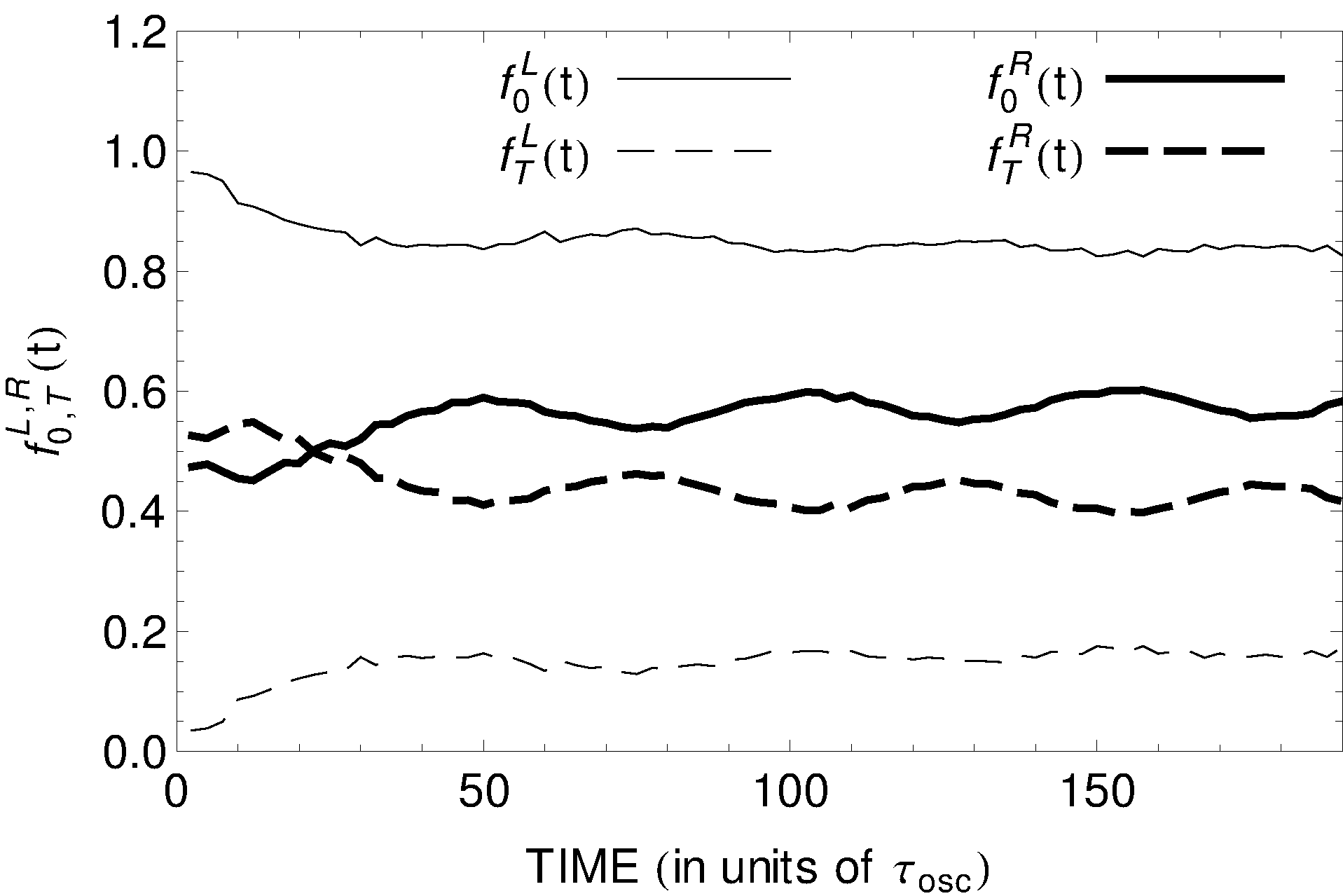}}
\resizebox{3.1in}{2.1in} {\includegraphics{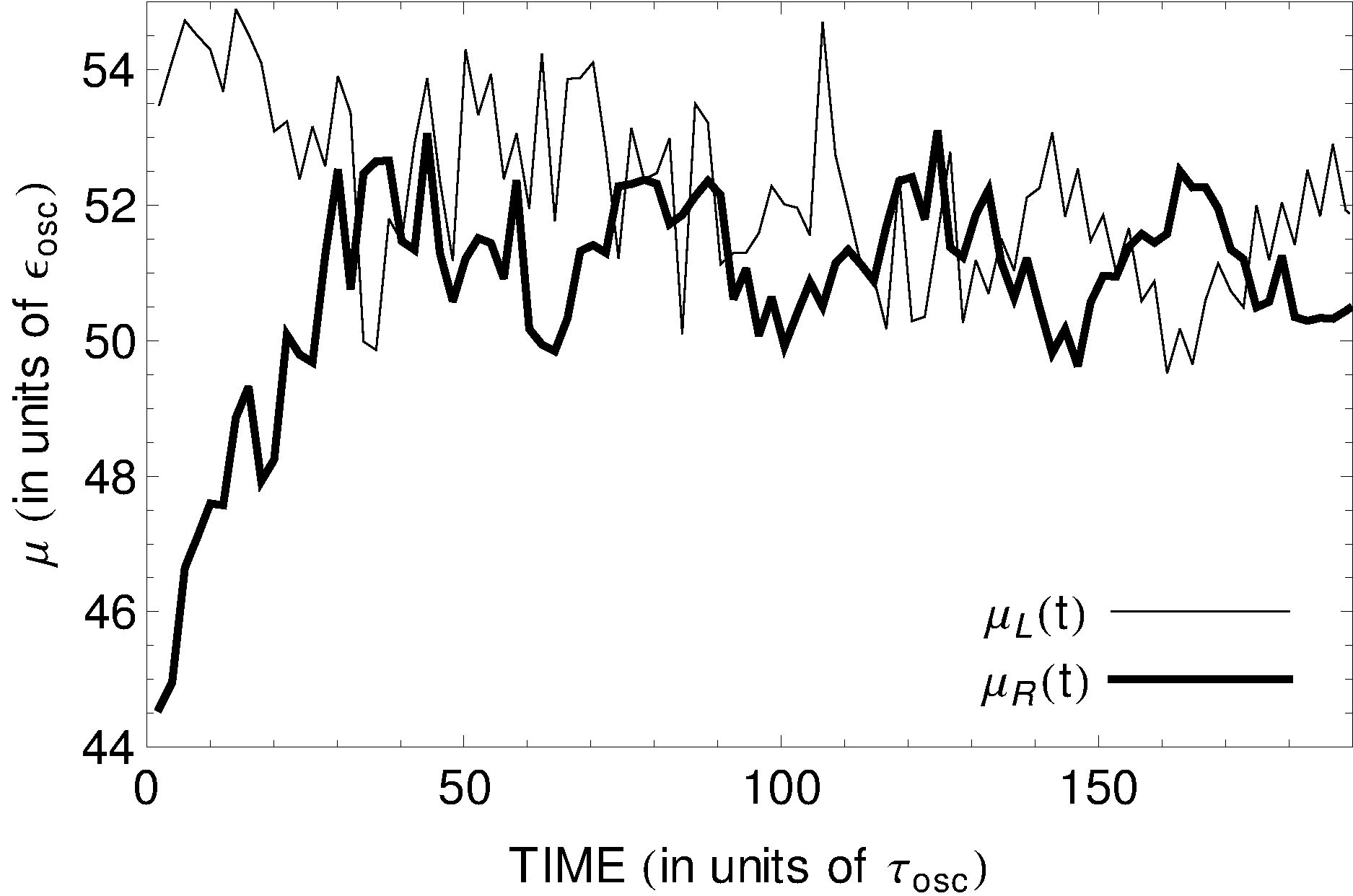}}
\caption{The upper frame shows the time evolution of the condensate (solid line) and of the thermal (dashed line) fractions in the left 
vessel (thin lines) and in the right vessel (thick lines). The time unit is $\tau_{{\rm osc}} = 6.366$ms. 
The initial condensate fraction in the right vessel is about $50$\% ($T=83$nK) and the channel 
width is $W_c =\ell_{{\rm osc}} \approx 4.2\mu$m . As one can see, the condensate and thermal fractions in each vessel never reach the same level meaning that the system does not reach thermal equilibrium. The lower frame shows the local chemical potentials calculated in the left (thin line) and in the right (thick line) vessels. As one can see, the system is able to reach rapidly, in about $31 \tau_{{\rm osc}}$ ($200$ms), a state very close to thermodynamical equilibrium ($\mu_L \sim \mu_R$). The two distinctive features of the helium fountain effect are thus recovered: because the system achieves local thermodynamical equilibrium, a temperature gradient is immediately compensated by a pressure difference, which generates a particle flow. 
}
\label{chem}
\end{figure}

The initial chemical potential difference implies a pressure difference and the existence of a particle flow when the channel is opened. To prove that the thermo-mechanical effect is indeed present in our system, we have to show that mechanical equilibrium is reached at once whereas thermal equilibrium is never reached during the considerably long computation time of our simulations.

To this end we first compute and compare the relative condensate and thermal fractions $f_0^X(t)$ and $f_T^X(t)$ in the left ($X=L$) and the right ($X=R$) vessels:
\begin{align}
f_0^X(t) = \frac{N^X_0(t)}{N_X(t)} & = \frac{n_0^X(t)}{n_X(t)}, \\ 
f_T^X(t)=\frac{N_T^X(t)}{N_X(t)} & = 1-f_0^X(t).
\end{align}
The upper frame of Fig.\ref{chem} shows these quantities for an initial right condensate fraction of $50$\% ($T=83$nK) and the thinest channel width considered here, i.e $W_c=\ell_{{\rm osc}}\approx 4.2\mu$m. It is clearly visible that after $157 \tau_{{\rm osc}}$ ($1$s), the condensate fraction in the left well is much larger than in the right well. This situation will hold obviously even longer. Similarly the thermal components in both vessels are very different. This signifies that both subsystems are not in thermal equilibrium. During the evolution, the initial hot cloud in the right vessel always remains much hotter then in the left part.

To show that the system (almost) reaches mechanical equilibrium after a short period of time, we have to consider the chemical potential 
defined according to (\ref{SCHF5}):
\begin{equation}
\mu({\bf r}) = g\, \rho_0({\bf r}) + 2\, g\, \rho_{T}({\bf r}) + V_{tr}({\bf r}) \,.
\label{CHPOT1}
\end{equation}
At mechanical equilibrium, the chemical potential should be position-independent. For comparison we choose two positions on opposite sides of the barrier located near the maximum of the initial atomic densities in each wells, ${\bf r}_R = (x,y,z)$ and ${\bf r}_L = (-x,y,z)$, and we calculate the corresponding local chemical potentials $\mu_L=\mu({\bf r}_L)$ and $\mu_R=\mu({\bf r}_R)$. There is however, one technical difficulty. The condensate and thermal densities are obtained from the diagonalization of the column-averaged one-particle density matrix. Therefore, in fact we only know the 2D densities in the ($Ox,Oy$) plane for all eigenmodes. To estimate the 3D densities, we need to calculate the $Oz$-width of each eigenmode along the channel. For this we average the one-particle density matrix Eq.\eqref{denmat} along $Oy$ ending up with column densities in the ($Ox,Oz$) plane. We extract the $Oz$-width $w_k^z(x)$ for each mode along the channel as the FWHM of the corresponding column densities. The 3D density is then estimated through $\rho_k(x,0,0) = \rho_k^{xy}(x,0)/w_k^z(x)$, where $\rho_k^{xy}$ is the column density of the k-th mode in the ($Ox,Oy$) plane.

\begin{figure}[htb]
\resizebox{3.4in}{2.1in} {\includegraphics{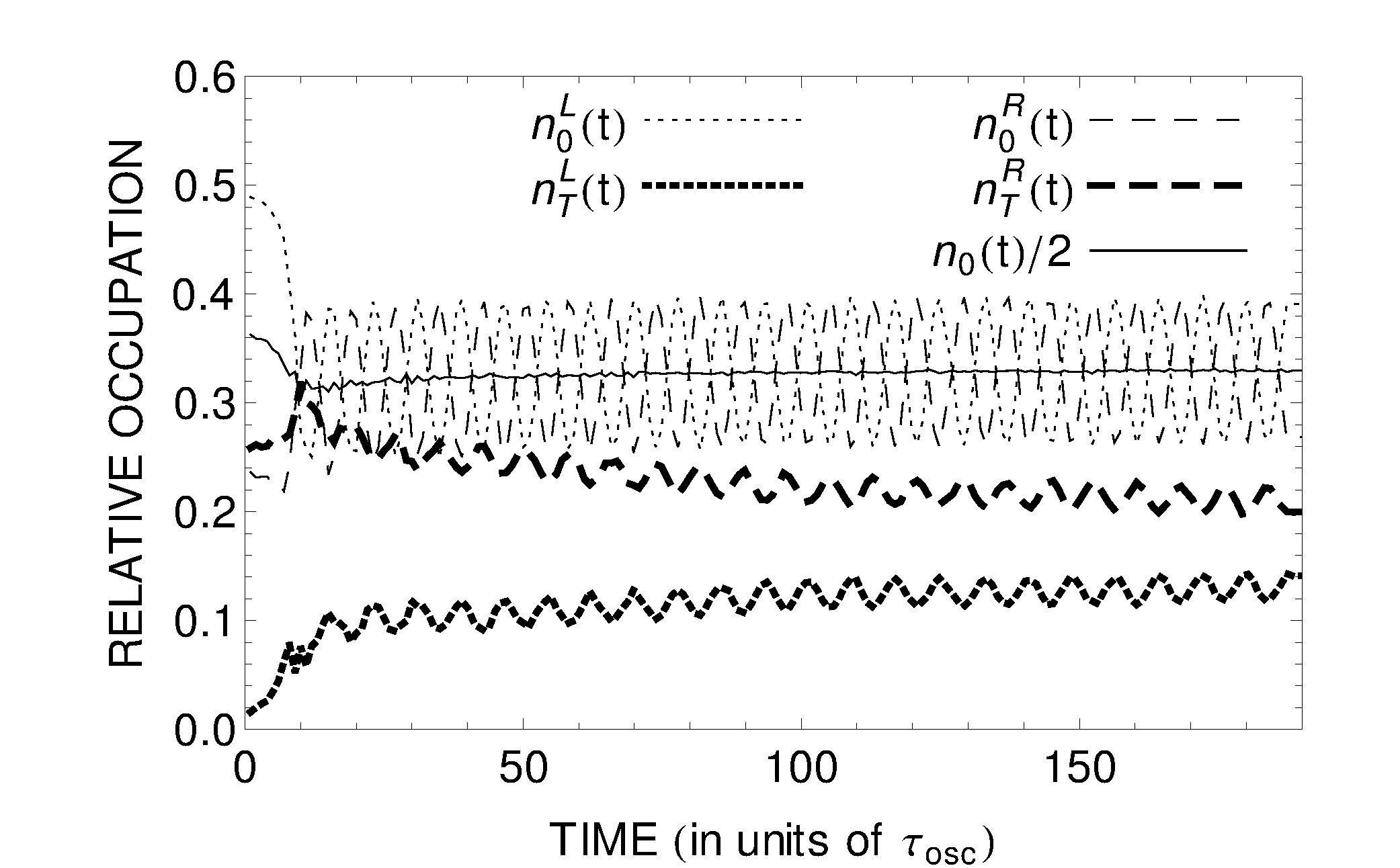}}
\caption{Time evolution of the relative occupation numbers of the condensate $n^{L,R}_0(t)$ (thin lines) and of the thermal cloud $n^{L,R}_T(t)$ (thick lines) in the left (dotted lines) and in the right (dashed lines) vessels. The time unit is $\tau_{{\rm osc}} = 6.366$ms. The initial condensate fraction in the right well is $50$\% ($T=83$nK) and the final width of the channel is $W_c=4.0 \ell_{{\rm osc}}$ ($16.8\mu$m).
As one can see, after a short initial stage, the right and left condensate fractions oscillate around a mean value which is half the total 
condensate fraction $n_0(t)/2$ (thin solid line). The thermal part, after a while, stays roughly constant but, 
as clearly seen, some part of the thermal cloud flows in phase with the condensed atoms. 
}
\label{relocppo3}
\end{figure}

Having the 3D densities, we can calculate the chemical potentials $\mu_L$ and $\mu_R$ as the average over a few points located around $x=-4.6$ and $x=4.6 \ell_{{\rm osc}}$ ($19.3\mu$m) respectively. The time evolution of these chemical potentials is shown in the lower frame of Fig.\ref{chem}. Although the curves look a bit ragged, we nevertheless see that the system rapidly reaches a state very close to mechanical equilibrium, $\mu_L \sim \mu_R$, in about $31 \tau_{{\rm osc}}$ ($200$ms). 
In fact we observe small out-of-phase oscillations of the chemical potentials caused by the back-and-forth oscillations of the condensed atoms.

As a main conclusion of the above discussions, we see that our system does present all the three distinctive features of the helium fountain experiment: (i) the system cannot achieve thermal equilibrium, (ii) 
a state oscillating slightly around mechanical equilibrium is reached, 
 and (iii) the component which flows through the very narrow channel connecting the two vessels at different temperatures does not transport heat.

\subsection{Superfluid and normal component}
\label{SN}

To show that our system was not reaching thermal equilibrium, we had to divide the classical field into a condensate and a thermal component. 
As the condensate component corresponds to the dominant eigenvalue of a coarse-grained one-particle density matrix, the thermal cloud consists of many modes with relatively small occupation numbers. This coarse-graining procedure splits the system into many different modes.
On the other hand the standard two-fluid model of the helium fountain is solely based on the distinction between a superfluid and a normal component. 
For liquid helium, which is a strongly interacting system, there is an essential difference between the condensate and the superfluid component. This difference is much less pronounced in the case of weakly-interacting trapped atomic condensates, but is nevertheless noticeable as pointed out in \cite{Zawitkowski} where the macroscopic excitation of a nonzero momentum mode has been studied within the classical fields formalism for a homogeneous box-like system.

As will be shown in this section, the situation is somehow similar for the inhomogeneous system studied here: our numerical results show that not only the lowest mode (the condensate part) of the coarse-grained one-body density matrix does contribute to the superfluid flow but also some excited modes.
A careful reader might have already noticed that in Fig.\ref{denpo} some part of the thermal component oscillates together with the condensate. This effect is very small for very narrow channels but is becoming quite pronounced for wider channels. Fig.\ref{relocppo3} shows the dynamics of the relative occupation numbers of the condensate and of the thermal components when the channel width is $W_c=4.0 \ell_{{\rm osc}}$ ($16.8\mu$m), the initial condensate fraction in the right vessel being $50$\%. It is clearly visible that a certain amount of excited atoms is flowing in phase with the condensate, back and forth from one vessel to the other. 

\begin{figure}[htb]
\resizebox{3.2in}{4.2in} {\includegraphics{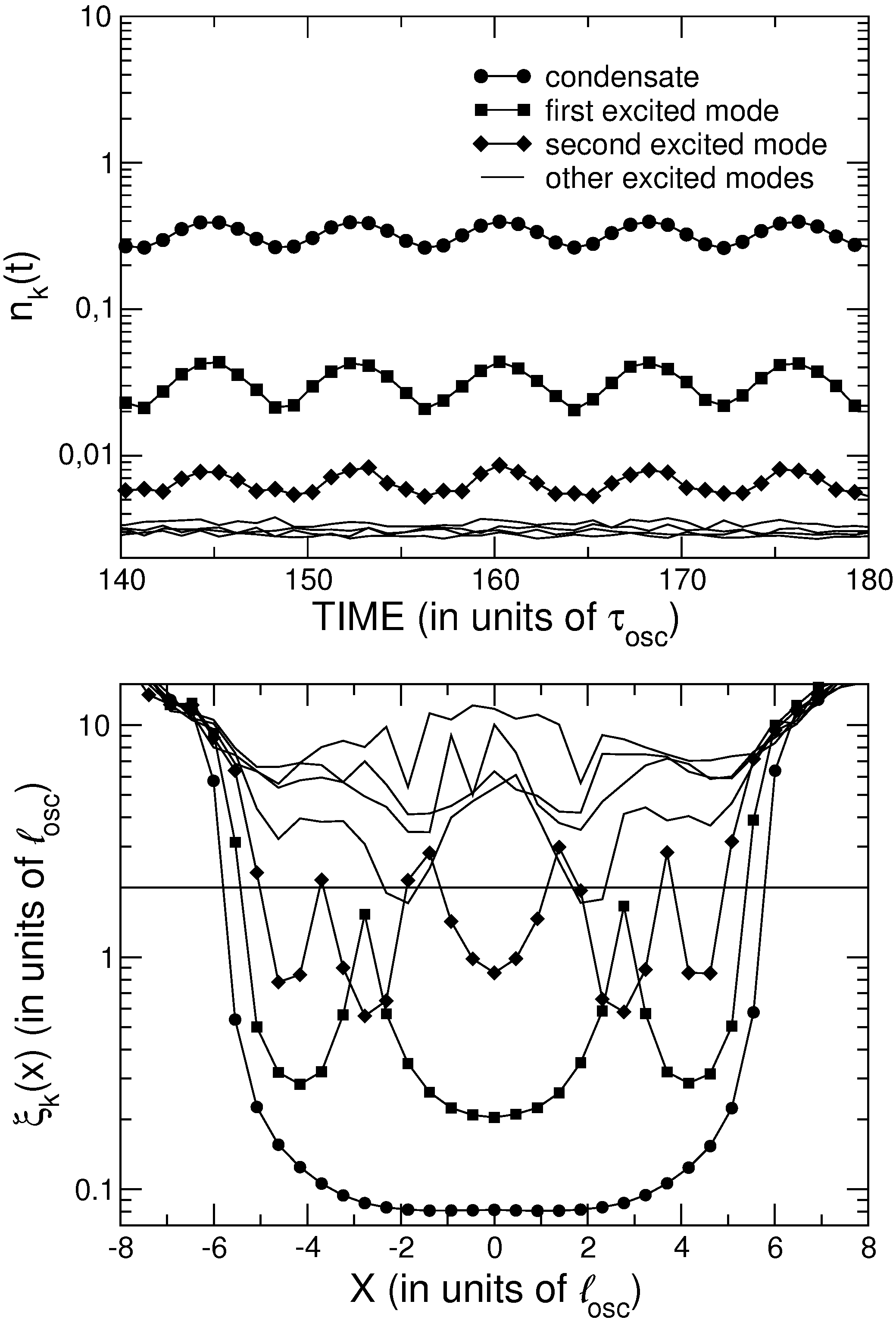}}
\caption{Time evolution of the relative occupation numbers of the seven dominant modes in the right vessel of the system (top frame). The time unit is $\tau_{{\rm osc}}$. The solid circles, solid squares, and solid diamonds correspond respectively to the condensate and to the next two highest occupied modes. The remaining four thin lines correspond to the next four modes of smaller occupation numbers. The bottom frame shows $\xi_k(x,0,0)$ for these modes. The thin horizontal line corresponds to the half the width of the channel, $W_c/2$. The initial condensate fraction in the right vessel is about $50$\% ($T=83$nK) and the channel width is $W_c=4.0 \ell_{{\rm osc}}$ ($16.8\mu$m). 
}
\label{norms}
\end{figure}

To explain this behavior, we show in Fig.\ref{norms} the time evolution of the relative occupation numbers of the first seven dominant eigenmodes (in the right well) of the one-particle density matrix, the largest occupation number corresponding to the condensate.
Apart from the condensate, the next two modes in the hierarchy exhibit very similar, fast and in-phase oscillations and 
their occupation numbers are significantly larger than those of the remaining other modes. These two modes, together with the condensate, constitute the three largest coherent `pieces' of the system. 

To tentatively explain why these three modes can flow freely from one vessel to the other, we define the following local lengths for each of these modes, $\xi_k(x,0,0)=1/\sqrt{8\pi a \rho_k(x,0,0)}$ where $x$ is a distance along the channel direction, $a$ is the scattering 
length and $\rho_k$ is the 3D density of the mode estimated through the CFA procedure described in the Appendix. These local lengths are shown in the bottom frame of Fig.\ref{norms}.  The thin horizontal line corresponds to half the width of the channel, $W_c/2$. We immediately see that the modes flowing together with the condensate fulfill the condition:
\begin{equation}
\label{XiW}
 \xi_k \lesssim \frac{W_c}{2},
 \end{equation}
where $\xi_k$ is the "typical" local length of mode $k$ (for example, taken at the middle of the channel).
As a rule of thumb, we infer that only modes with a typical local length smaller than half the channel width can flow freely. It is worth mentioning that criterion Eq.\eqref{XiW} somehow interpolates between the superfluid strong link (for which $\xi_k \ll W_c$) and weak link (for which $W_c\lesssim \xi_k$) regimes \cite{WeakLinks}. The modes satisfying $ \xi_k \lesssim \frac{W_c}{2}$, condensate included, seem to form the superfluid component. Higher modes, having a typical local length larger than $W_c/2$, cannot fit into the channel and cannot flow: they seem to form the normal component. A careful reader might have noticed that these local lengths look like healing lengths associated to each of the excited modes. In that respect, associating a local length to each excited mode of the one-body density matrix might seem dubious but could maybe be justified in the sense that the Onsager-Penrose criterion works ideally in the thermodynamic limit. Here we deal with a finite-size system where the observed differences between the fractions extracted from the one-body density matrix are maybe not enough macroscopic (typically we get a factor of 15 between the ground state and first excited mode and a factor 3 between the first and second excited mode, see Fig.\ref{norms}).
In this case one maybe faces a situation somehow similar to fragmentation \cite{EJMueller, Fallani}, the elongated (quasi-1D) nature of the channel favoring the emergence of different fluids flowing in the pipe. We plan to investigate this problem in a future publication.
  
Applying Eq.\eqref{XiW} as a rule of thumb, the superfluid fraction and superfluid density are then respectively defined as
\begin{eqnarray}
&n_S(t) = \sum_{k=0}^{k_S} n_k(t), \nonumber \\
&\rho_S(x,y,t) =  \sum_{k=0}^{k_S}|\psi_k(x,y,t)|^2,
\end{eqnarray}
where $k_S$ is the index of the highest occupied one-particle density matrix eigenmode fulfilling $\xi_k<W_c/2$. Analogously one can define corresponding quantities for the normal component, i.e. $n_N(t)$ and $\rho_N(x,y,t)$. It is moreover convenient to split the superfluid and normal fractions into their left and right components $n_{S,N}^{L,R}(t)$.

\begin{figure}[htb]
\resizebox{3.4in}{2.1in} {\includegraphics{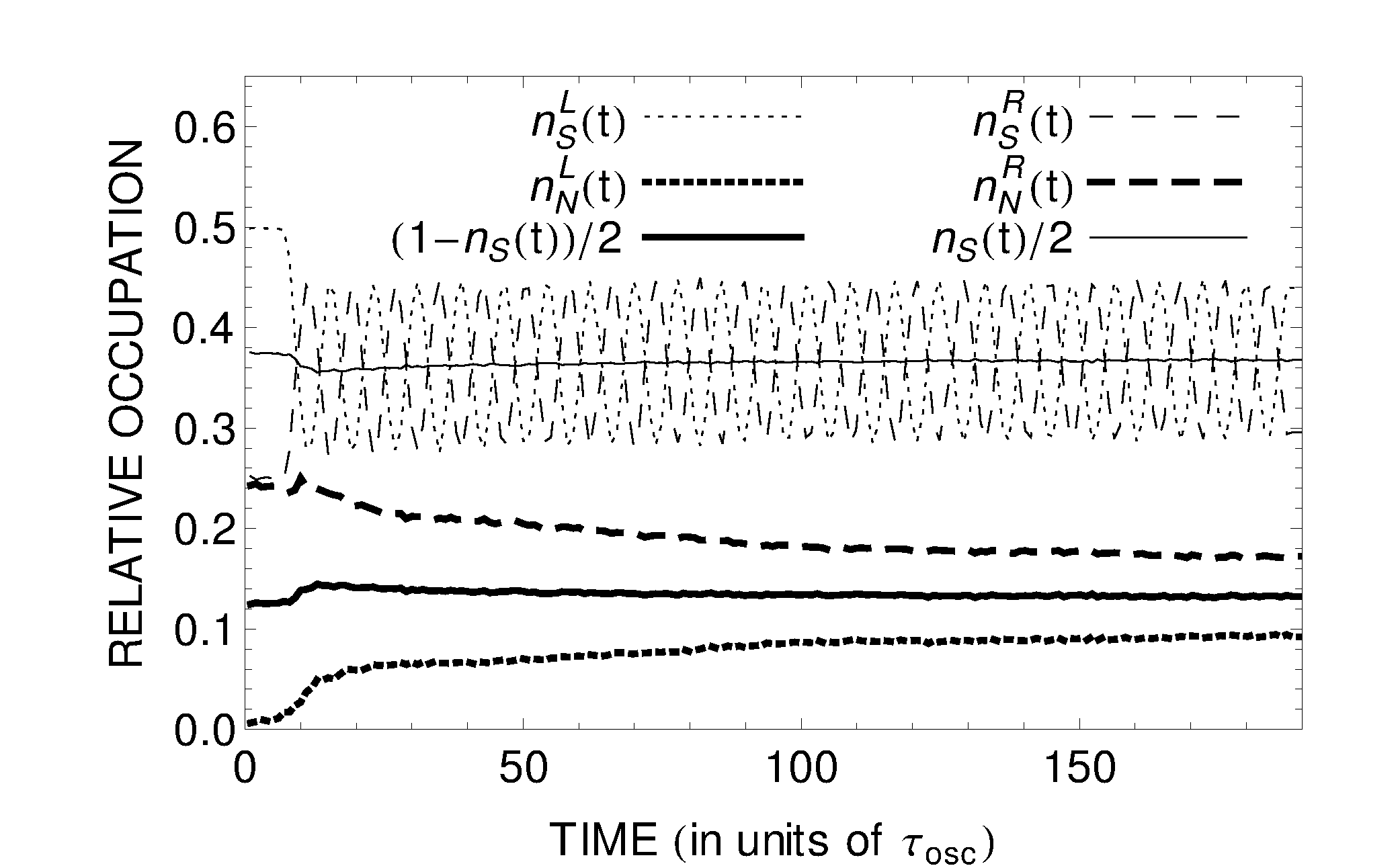}}
\caption{
Time evolution of the relative occupation numbers of the superfluid (thin lines) and normal (thick lines) components in the left (dotted lines) and right (dashed lines) vessels. The time unit is $\tau_{{\rm osc}}$. The system contains initially $50$\% ($T=83$nK) of condensed atoms in the right potential well and the final width of the channel is $W_c=4.0 \ell_{{\rm osc}}$ ($16.8\mu$m). The thin and thick solid lines show half the total superfluid and normal fractions respectively. The normal fraction flows smoothly and slowly from the hotter vessel to the colder one as expected while the superfluid fraction oscillates back and forth between the two vessels around a mean value being half the total superfluid fraction $n_S(t)/2$ (thin solid line). The thick solid line represents half the total normal fraction $(1-n_S(t))/2$ which is never reached by the left and right normal components during the time scale of the simulation.
}
\label{relocpsp2}
\end{figure}

\begin{figure}[htb]
\resizebox{3.in}{5.81in} {\includegraphics{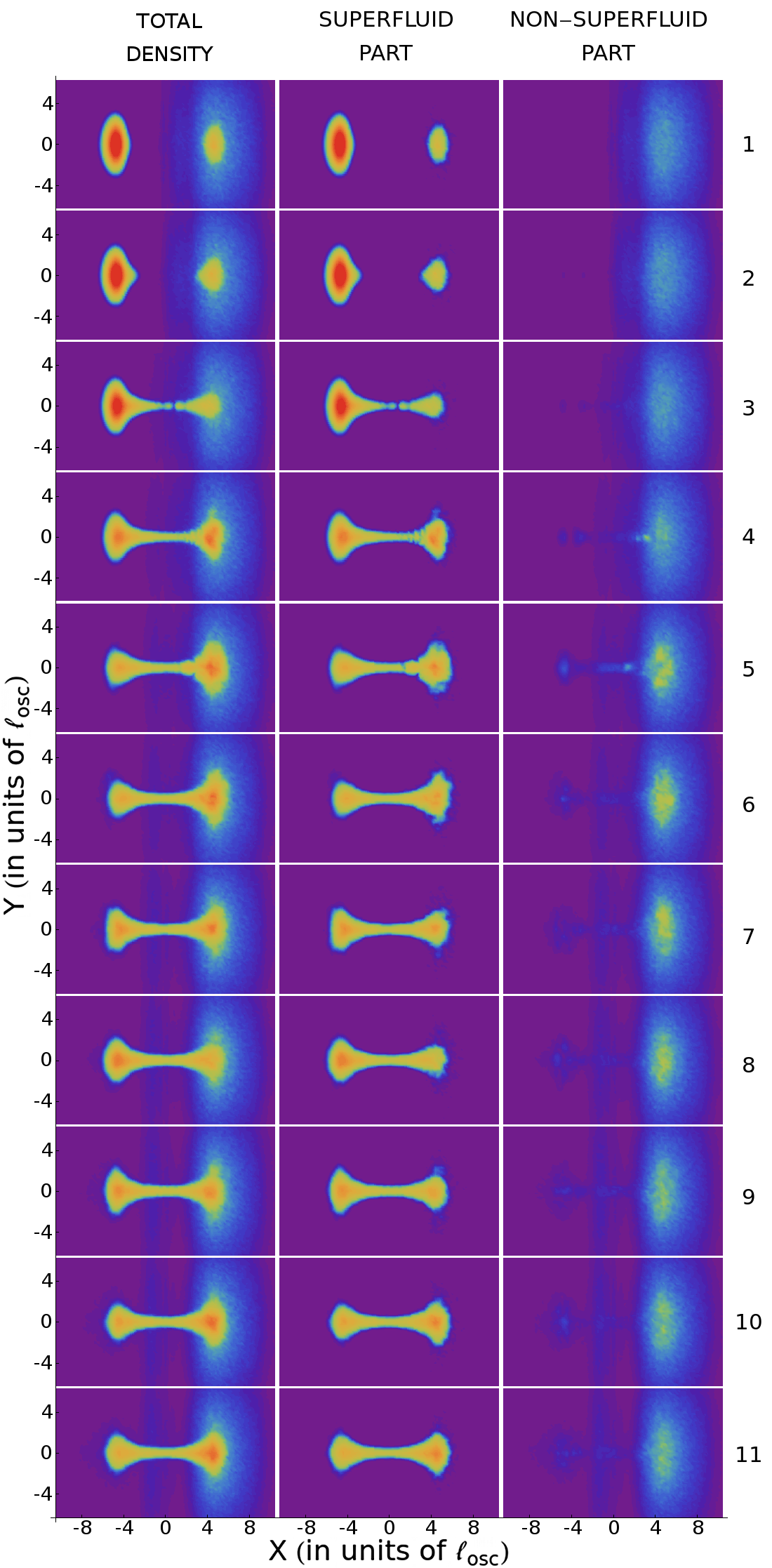}}
\caption{(Color online). Snapshots of the time evolution of the total (left), superfluid (middle) and normal (right) column densities.
The initial condensate fractions are $100$\% ($T=0$) in the left vessel and about $20$\% ($T=100$nK) in the right vessel. The final channel width is $W_c = 2.4 \ell_{{\rm osc}}$ ($10\mu$m). The time interval between the frames is about $2.5 \tau_{{\rm osc}}$ ($15.9$ms). As clearly seen, the superfluid component oscillates back and forth between the two vessels while the normal component is essentially trapped in the hotter right vessel.}
\label{densp}
\end{figure}

These quantities are shown in Fig.\ref{relocpsp2}. It can be seen that the normal component flows only very slowly from the hotter to the colder well as it is expected for the superfluid fountain effect. Comparison with Fig.\ref{relocppo3} further shows that the normal component, contrary to the thermal one, does not exhibit any temporal oscillations. 
In Fig.\ref{densp}, we plot the superfluid column density $\rho_S(x,y,t)$ (middle column), the normal column density 
$\rho_N(x,y,t)$ (right column), and the total atomic density $\bar{\rho}(x,y,x,y;t)$ (left column) 
for the same parameters as in Fig.\ref{denpo}. The left column is identical as in fig. \ref{denpo} and is put here as a reference. 
One can clearly see that essentially only the superfluid component travels back and forth between the two vessels. The normal component remains mainly located in the right hotter vessel and its flow to the colder left vessel is almost invisible. The similarity between the AC-Josephson effect obtained in weak links and BEC experiments \cite{WeakLinks, AC-Jo1, AC-Jo2} and the temporal oscillatory behavior of the superfluid component here is appealing. However, our system is more complex and would require a deeper analysis to substantiate this similarity in a quantitative way. For example, in our case, the chemical potential difference displays some (noisy) temporal oscillations and their link to the period of ocillation of the superfluid component remains to be made. We will address these points in a future work.

To estimate the rate of flow of the superfluid fraction, we wait for the system to reach its oscillatory regime and then fit the (damped) oscillations of the superfluid fraction in the left vessel by:
\begin{equation}
 F(t)=A\sin{(2\pi\nu t+\phi)} e^{-\gamma t} + B t + C,
\end{equation}
and extract the oscillation frequency $\nu$ and
the oscillation amplitude $A$ of the superfluid flow. The maximal superfluid flux through the channel is $F_S = 2\pi A N \nu$. We also fit the slow decrease of the normal fraction in the left vessel by the linear function $G(t)=\alpha t +\beta$. 
The maximal flux of the normal atoms is then $F_N = \alpha N$. All these quantities are collected in Tables  \ref{fity20} and \ref{fity50}.
\begin{table}[htb]
\begin{tabular}{|c||c|c|c|c|c|c|}
\hline
 $W_c$[osc.u.]	&$A$	&$\nu$[Hz]	&$\alpha$[s$^{-1}$]	       &$F_S$[$\frac{at}{ms}$]        &$F_N$[$\frac{at}{ms}$]	\\
\hline\hline
0.96		&0.0254	&3.82		&0.0201		         &152                  &5.03		\\
1.2		&0.036	&5.27		&0.0277		         &298                  &6.93		\\
2.4		&0.06	&12.6		&0.0502		         &1188                &12.55		\\
\hline
\end{tabular}
\caption{The relevant coefficients obtained from our fitting procedure and the calculated superfluid and normal rates of flow. The initial occupation number of the condensate in the right well is $20$\% ($T=100$nK). 
}
\label{fity20}
\end{table}

\begin{table}[htb]
\begin{tabular}{|c||c|c|c|c|c|c|}
\hline
 $W_c$[o.u.]	&$A$	&$\nu$[Hz]	&$\alpha$         &$F_S$[$\frac{at}{ms}$]	   &$F_N$[$\frac{at}{ms}$]	\\
\hline\hline
1.0		&0.036	&3.8		&0.013		         &215                  &3.25		\\
2.0		&0.078	&9.1		&0.027		         &1150                &6.75		\\
4.0		&0.085	&20.1		&-		         &2684                 &-		\\
\hline
\end{tabular}
\caption{The relevant coefficients obtained from our fitting procedure and the calculated superfluid and normal rates of flow. The initial occupation number of the condensate in the right well is $50$\% ($T=83$nK). 
}
\label{fity50}
\end{table}
Note, that the last row of Table \ref{fity50} does not contain any value for the $\alpha$ coefficient
nor for the corresponding normal flux $F_N$. This is because, for wider channels, the rate of flow of the normal component is changing significantly in time and fitting the decrease by a linear function is no longer reasonable. In this case, the flow is fastest at the beginning as it is visible in Fig.\ref{relocpsp2}. 

We did not include the value of the coefficients $B$, $C$, and $\beta$ in the Tables, even if they increase the precision of our  fitting procedure, as they are essentially irrelevant four our considerations. For channel widths $W_c \leq 5\ell_{{\rm osc}}$, the coefficients $\gamma$ turns out to be smaller than the statistical error ($\gamma \sim 0$) and are also not included in Tables \ref{fity20} and \ref{fity50}. This observation is in agreement with the fact that the dynamics takes place in the collisionless regime as mentioned in the Introduction.

We see that both superfluid and normal flow rates increase with the channel width. Moreover, the superfluid flow rate is in all cases larger by two or three orders of magnitude then the normal one. We expect that the normal component behaves like a classical fluid. Therefore, its flow rate should correspond to the flux of atoms distributed initially according to the classical phase space distribution as obtained from the SCHFM equations described in the Appendix. Our SCHFM calculations indeed give a value very close to the one obtained from the classical fields dynamics. For example the flux of thermal atoms for a system initially prepared with $50$\% of condensed atoms in the hotter vessel and for a final channel width $W_c = 6.0 \ell_{{\rm osc}}$ ($25.2\mu$m) is found to be $F_N \approx 115.4$ atoms/ms. The classical field approximation gives a similar result $F_N \approx 182.5$ atoms/ms. Indeed, the very slow transfer of the normal component is a phase space distribution effect -- a very small fraction of thermal atoms have velocities aligned along the channel. On the contrary, the superfluid 
component is built from coherent modes. The coherence of these modes extends over the entire two vessels and is established on a short time scale of about $16 \tau_{{\rm osc}}$ ($100$ms). 

\section{Conclusions}
\label{concl}
In conclusion, we have shown that the analog of the thermo-mechanical effect, observed in the celebrated superfluid helium II fountain, could be also observed with present-day experiments using weakly-interacting degenerate trapped alkali gases. We have proposed a realistic experimental setup based on a standard harmonic confinement potential and analyzed it with the help of the classical fields aproximation method. The trapped ultracold gas is first split in two subsytems thanks to a potential barrier. Each of the two independent subsystems achieve their own thermal equilibrium, the final temperature in the two vessels being different. At a later time, a communication channel is opened between the two vessels, and the atoms are allowed to flow from one vessel to the other. We have shown that the transport of atoms between the two subsystems prepared at two different temperatures exhibits the three main features of the superfluid fountain effect: the thermodynamical equilibrium is obtained almost instantly while the thermal equilibrium is never reached, implying in turn a pressure difference and a superfluid flow. Our numerical data seem also to show that the superfluid component of this system is composed of all eigenmodes of the one-particle density matrix having a sufficiently small healing length that can fit into the communication channel. The superfluid flow is at least two orders of magnitude faster than the flow of the normal component. The slow flow of the normal component can be understood as a phase space effect. As we pointed out, our trap design and the long-time temporal oscillations of the superfluid component bear similarities with superfluid weak links and the associated AC-Josephson effect. Future studies aim at investigating this similarity and to further understand the validity of the criterion Eq.\eqref{XiW}.
\vspace{0.3cm}

\acknowledgments

The Authors wish to thank Miros\l{}aw Brewczyk, Bj\"orn Hessmo, Cord M\"uller and David Wilkowski for discussions and valuable comments. Special thanks go to Nicolas Pavloff for pointing to us the similarity between our experimental design and superfluid Josephson weak and strong links and to Iacopo Carusotto for a critical reading of our manuscript.
TK and MG acknowledge support from the Polish Goverment research funds for the period 2009-2011 under the grant N N202 104136. MG acknowledges support from EU STREP NAME-QUAM. ChM and BG acknowledge support from the CNRS-CQT LIA FSQL and from the France-Singapore Merlion program, FermiCold grant No. 2.01.09. The Centre for Quantum Technologies is a Research Centre of Excellence funded by the Ministry of Education and the National Research Foundation of Singapore.

\appendix
\section{Classical fields approximation}
\label{method}

There are different effective methods to describe and study dynamical effects in condensates at nonzero temperature. For example, the Zaremba-Nikuni-Griffin formalism assumes a splitting 
of the system into a condensate and a thermal cloud \cite{Zaremba} whereas different versions of the classical fields method describe both the condensate and the thermal cloud by a single Gross-Pitaevskii equation  \cite{Goral,Davis,Sinatra,Svistunov,Berloff}. Here, we will use the classical fields method as described in \cite{brew1} and optimized in \cite{brew2} for arbitrary trapping potentials.
To be self-contained, this Appendix gives the rationale of the CFA method and how it is applied to our system.

\subsection{Formalism}
\label{formalism}

We start with the usual bosonic field operator $\hat {\Psi }({\bf r},t)$ which annihilates 
an atom at point ${\bf r}$ and time $t$. It obeys the standard bosonic commutation relations:
\begin{align}
 \left[ {\hat {\Psi }({\rm {\bf r}},t),\hat {\Psi }^+({\rm {\bf r}'},t)}
\right] & =  \delta ({\rm {\bf r}}-{\rm {\bf r}'}) \nonumber \\
 [\hat {\Psi}^+({\rm {\bf r}},t),\hat {\Psi}^+({\rm {\bf r}'},t)] & =  0 \nonumber \\
 [\hat {\Psi}({\rm {\bf r}},t),\hat {\Psi}({\rm {\bf r}'},t)] & =  0,
\label{comrel}
\end{align}
and evolves according to the Heisenberg equation of motion:
\begin{eqnarray}
&&i\hbar \frac{\partial}{\partial t} \hat {\Psi }({\rm {\bf r}},t) =
\left[ -\frac{\hbar^2}{2m} \nabla^2 + V_{tr}({\rm {\bf r}},t)    \right]
\hat {\Psi }({\rm {\bf r}},t)   \nonumber  \\
&&+ g\, \hat{\Psi }^+({\rm {\bf r}},t) \hat {\Psi }({\rm {\bf r}},t)
\hat {\Psi }({\rm {\bf r}},t)   \,,
\label{Heisenberg}
\end{eqnarray}
where $V_{tr}({\rm {\bf r}},t)$ is the (possibly) time-dependent trapping potential and $g=4\pi \hbar^2 a /m$ is the coupling constant expressed in terms of the s-wave scattering length $a$.

The field operator itself can be expanded in a basis of one-particle
wave functions $\phi_\alpha({\bf r})$, where $\alpha$ denotes the set of all necessary one-particle quantum numbers:
\begin{equation}
\hat {\Psi }({\rm {\bf r}},t) = \sum_\alpha \phi_\alpha({\bf r})  \hat {a}_\alpha(t).
\label{expansion}
\end{equation}
In the presence of a trap, a natural choice for the one-particle modes $\phi_\alpha$ would be the harmonic oscillator modes, otherwise one generally uses plane wave states. The classical fields method is an extension of the Bogoliubov idea to finite temperatures and gives some microscopic basis to the two-fluid model. The main idea is to assume that modes $\phi_\alpha$ in expansion (\ref{expansion}) having an energy $E_\alpha$ less than a certain cut-off energy $E_c$ are macroscopically occupied and, consequently, to replace all corresponding annihilation operators by $c$-number amplitudes:
\begin{equation}
\hat {\Psi }({\rm {\bf r}},t) \simeq \sum_{E_\alpha\leq E_c} \phi_\alpha({\bf r})  a_\alpha(t) + \sum_{E_\alpha > E_c} \phi_\alpha({\bf r})  \hat {a}_\alpha(t) \;.
\label{expansion2}
\end{equation}
Assuming further that the second sum in (\ref{expansion2}) is small and can be neglected, 
the field operator $\hat {\Psi }({\bf r},t)$ is turned into a classical complex wave function:
\begin{equation}
\hat {\Psi }({\rm {\bf r}},t) \to \Psi ({\rm {\bf r}},t) = \sum_{E_\alpha \leq E_c} \phi_\alpha({\bf r}) a_\alpha(t)  \,.
\label{expansion1}
\end{equation}
In this way, both the condensate and a thermal cloud of atoms, interacting with each other, will be described by a single classical field $\Psi({\bf r},t)$. Injecting (\ref{expansion1}) into (\ref{Heisenberg}), we obtain the equation of motion for
the classical field:
\begin{eqnarray}
&&i\hbar \frac{\partial}{\partial t} {\Psi }({\rm {\bf r}},t) =
\left[ -\frac{\hbar^2}{2m} \nabla^2 + V_{tr}({\rm {\bf r}},t)    \right]
{\Psi }({\rm {\bf r}},t)   \nonumber  \\
&&+ g\, {\Psi }^*({\rm {\bf r}},t) {\Psi }({\rm {\bf r}},t)
{\Psi }({\rm {\bf r}},t)   \,.
\label{CFequation}
\end{eqnarray}
In numerical implementations, one controls a total energy, a number of macroscopically occupied modes $\phi_\alpha$ and a value of $gN$. The energy-truncation constraint $E_\alpha\leq E_c$ is usually implemented by solving Eq. (\ref{CFequation})
on a rectangular grid using the Fast Fourier Transform technique. The spatial grid step
determines the maximal momentum per particle, and hence the energy, in the system whereas the
use of the Fourier transform implies projection in momentum space.

Equation (\ref{CFequation}) looks identical to the usual Gross-Pitaevskii equation
describing a Bose-Einstein condensate at zero temperature. However, the interpretation
of the complex wave function $\Psi ({\bf r},t)$ is here different. It describes {\it all} the atoms
in the system. Therefore, the question arises on how to extract all these modes out of the time-evolving classical field $\Psi ({\bf r},t)$. For this purpose, we follow the definition of Bose-Einstein condensation originally proposed by Penrose and Onsager \cite{POdef} where the condensate is assigned
to be described by the eigenvector corresponding to the dominant eigenvalue of the one-particle density matrix. However, it has been noticed in \cite{Pethick00} that the Penrose-Onsager criterion has to be modified when the system moves with an amplitude larger than its size. To get the condensate fraction, defined as the largest coherent contribution to the many-body wave function, one should perform measurements in the center of mass reference frame. This corresponds to a joint simultaneous detection
of many (several at least) particles \cite{Gajda}. In the mean-field approach, such a definition leads to a dominant eigenmode of the instantaneous one-particle density matrix. This one-particle density matrix reads:
\begin{equation}
\varrho^{(1)}({\bf r},{\bf r}^{\,\prime};t) = \frac{1}{N}\, \Psi({\bf r},t)\, 
\Psi^*({\bf r}^{\,\prime},t)   \,,
\label{denmat}
\end{equation}
and obviously corresponds to a pure state with all atoms in the condensate mode. This is because Eq. (\ref{denmat}) 
is the spectral decomposition of the one-particle density matrix. To extract the modes out of the classical field some kind of averaging of the instantaneous one-particle density matrix is needed. 

\subsection{Coarse-grained one-body density matrix}

Recalling that in a typical experiment, one generally measures the column density integrated along some direction, we will implement here the same type of procedure and define the instantaneous coarse-grained density matrix:
\begin{equation}
\bar{\varrho}(x,y,x',y';t) = \frac{1}{N}  \int dz \, \Psi(x,y,z,t) \, \Psi^*(x',y',z,t)  \,,
\label{rhoave}
\end{equation}
from which we extract the corresponding eigenvalues in order to apply the Penrose-Onsager criterion \cite{Remark}.

Solving the eigenvalue problem for the coarse-grained density matrix (\ref{rhoave}) leads to
the decomposition:
\begin{equation}
\bar{\varrho}(x,y,x',y';t)=\sum_{k=0}^{K} n_k(t) \, \varphi_k(x,y,t) \, \varphi^*_k(x^{\,\prime},y^{\,\prime},t)  \,,
\end{equation}
where the relative occupation numbers $n_k(t)= N_k(t)/N$ of the orthonormal macroscopically occupied modes $\varphi_k$ are ordered according to $n_0(t) \geq n_1(t) \geq (\hdots)\geq n_K(t)$. For future convenience, we define the eigenmodes of the coarse-grained one-particle density matrix which are normalized to the relative occupation numbers of these modes and the corresponding one-particle density matrix $\bar{\rho}_k$:
\begin{eqnarray}
\psi_k(x,y,t) =  \sqrt{\frac{N_k}{N}}\, \varphi_k(x,y,t)   \,, \nonumber \\
\bar{\varrho}_k(x,y,x',y';t) = \psi_k(x,y,t)\psi^*_k(x',y',t) \, ,
\label{funk}
\end{eqnarray}
such that $\bar{\varrho} = \sum_{k=0}^{K} \bar{\varrho}_k$ and $\varrho_T = \sum_{k=1}^{K} \bar{\varrho}_k$, the condensate being described by $\bar{\varrho}_0$.

According to the standard definition, the condensate wave function corresponds to $k=0$ and the thermal density is simply:
\begin{equation}
\rho_T(x,y,t) = \bar{\varrho}(x,y,x,y;t) - |\psi_0(x,y,t)|^2  \,.
\end{equation}
In an equilibrium situation, the relative occupation numbers $n_k$ do not depend on time. In this case, the total number of atoms is determined from the smallest eigenvalue of the one-particle density matrix through $n_K N= n_{cut}$, where $n_{cut}\approx 0.46$ for the 3D harmonic oscillator \cite{Witkowska}, from which one can infer the value of the interaction strength $g$. The temperature $T$ of the system is then given by the energy of this highest occupied mode. In out-of-equilibrium situations, the Onsager-Penrose criterion remains perfectly well defined. The relative occupation numbers $n_k$ will depend on time but one can always diagonalize the coarse-grained one-particle density operator and possibly find one extensive eigenvalue \cite{Leggett}. We will use this criterion to define the condensate fraction throughout our paper.

Let us note that any initial state evolving with the Gross-Pitaevskii equation reaches a state of thermal equilibrium characterized by a temperature, a total number of particles and an interaction strength $g$. However to obtain an equilibrium classical field for a given set of parameters, one has to properly choose the energy of the initial state and the cut-off parameter $E_c$. This task is time consuming because the temperature and the number of particles can only be assigned to the field {\it after} the equilibrium is reached. To speed up the preparation of the initial equilibrium state, we first solve the self-consistent Hartree-Fock model (SCHFM) \cite{Pethick}. This allows us to estimate quite accurately the energy of the state for a given temperature and particle number. The detailed description of this procedure can be found in \cite{brew2}. The SCHFM equations read:
\begin{eqnarray}
&&\rho_0({\bf r}) = \frac{1}{g}\left[ \mu - V_{tr}({\bf r}) - 2\, g\, \rho_{T}({\bf r}) \right]
\label{SCHF1}  \\
&&f({\bf r},{\bf p}) = \left( e^{ [{\bf p}^2/2m + V_{e}({\bf r}) - \mu ] / k_B T} -1 \right)^{-1}
\label{SCHF2}  \\
&&\rho_{T}({\bf r}) = \frac{1}{\lambda_T^3} \;\; 
g_{3/2}\left( e^{\left[\mu - V_{e}({\bf r}) \right] /k_B T } \right )
\label{SCHF3}  \\
&&V_{e}({\bf r}) = V_{tr}({\bf r}) + 2\,  g\,  \rho_0({\bf r}) + 2\,  g\,  \rho_{T}({\bf r})
\label{SCHF4}
\\
&&\mu = g\, \rho_0(0) + 2\, g\, \rho_{T}(0) + V_{tr}(0) \,,
\label{SCHF5}
\end{eqnarray}
where
\begin{equation}
\lambda_T = \frac{h}{\sqrt{2\pi mk_B T}}
\end{equation}
is the thermal de Broglie wavelength. The $g_{3/2}(z)$ function is given by the expansion:
\begin{equation}
g_{3/2}(z) = \sum_{n=1}^\infty \frac{z^n}{n^{3/2}}  \,.
\end{equation}

The main variables in this approach are the condensate density $\rho_0({\bf r})$ and the phase space distribution function $f({\bf r},{\bf p})$ of thermal component. The thermal density $\rho_{T}({\bf r})$ can be obtained from $f({\bf r},{\bf p})$ by integrating over momenta. The effective potential $V_{e}({\bf r})$ and the chemical potential $\mu$ are functions of the condensate density and of the thermal density. The condensate and thermal densities can be found iteratively for a given number of atoms and condensate fraction by taking into account that the total number of atoms is $N=\int d{\bf r}\, (\rho_0({\bf r}) + \rho_{T}({\bf r}))$. The SCHFM is known to work well for the isotropic harmonic trap \cite{Krauth} and for inhomogeneous traps with small aspect ratio \cite{Aspect}. 
In the present work we also use the SCHFM equations
to extract the chemical potential and the thermal atoms distribution function in Section \ref{results}.


\begin{thebibliography}{99}
\bibitem{Remark} As a side remark, in close-to-equilibrium situations, the spatial average has been tested against a time average in \cite{brew2} for a classical field at thermal equilibrium in a harmonic trap: with a $1$\% accuracy, both methods gave the same result.
\end{thebibliography}
\end{document}